\documentclass[11pt]{article}
\usepackage{amstext,amsmath,amssymb,amsfonts,bbm}

\usepackage[utf8]{inputenc}
\usepackage{epsfig}
\usepackage{dsfont}

\usepackage[backend=biber, doi=true, url=true, style = numeric-comp, sorting = none]{biblatex}
\usepackage{amsthm}
\usepackage{subfigure}
\usepackage{color}
\usepackage{multirow}
\usepackage{psfrag}
\usepackage{graphicx}
\usepackage{extarrows}
\usepackage{authblk}
\usepackage[table]{xcolor}
\usepackage{microtype,booktabs,longtable,array,enumitem}
\usepackage[letterpaper,textwidth=17cm,textheight=23cm,centering]{geometry}
\usepackage{parskip}
\usepackage{physics}
\usepackage{comment}
\usepackage{stmaryrd}
\usepackage{mathbbol}
\usepackage{empheq}
\usepackage{tikz-cd}
\usepackage{tikz}
\usetikzlibrary{arrows.meta,calc,positioning,decorations.markings}
\usepackage{caption}
\usepackage{needspace}
\definecolor{ribbonblue}{RGB}{35,101,139}
\definecolor{facegreen}{RGB}{47,117,83}
\definecolor{cutred}{RGB}{157,61,55}
\definecolor{jtSage}{HTML}{E5EFE9}
\definecolor{jtBlue}{HTML}{E5EEF4}
\definecolor{jtGrey}{HTML}{F1F1F1}
\definecolor{jtRose}{HTML}{EDC3B5}
\definecolor{jtGold}{HTML}{F0D8A8}
\definecolor{jtInkBlue}{HTML}{47718D}
\definecolor{jtInkGreen}{HTML}{4D7965}
\tikzset{corner/.style={circle,fill=black,inner sep=1.7pt},
 transport/.style={thick,postaction={decorate},decoration={markings,
 mark=at position .55 with {\arrow{Latex}}}},
 treeedge/.style={thick,ribbonblue,-{Latex[length=2mm]}},
 freeside/.style={thick,dashed,cutred,-{Latex[length=2mm]}}}
\tikzset{
  jt edge/.style={draw=black!90,line width=.72pt,line cap=round,line join=round},
  jt faint/.style={draw=black!25,line width=.45pt,dash pattern=on 2pt off 2pt},
  jt arrow/.style={postaction={decorate},decoration={markings,mark=at position .55 with {\arrow{Stealth[length=1.65mm,width=1.15mm]}}}},
  jt tree/.style={draw=jtInkBlue,line width=1.35pt,line cap=round,line join=round},
  jt label/.style={font=\small,inner sep=1.4pt},
  jt move/.style={draw=black!75,line width=.8pt,-{Stealth[length=2.2mm,width=1.5mm]}}
}
\definecolor{ink}{RGB}{0,0,0}
\definecolor{sage}{RGB}{228,239,230}
\definecolor{rose}{RGB}{246,228,222}
\tikzset{edge/.style={thick,draw=ink},
 vertex/.style={circle,fill=ink,inner sep=1.6pt},
 move/.style={<->,>=Stealth,thick,draw=ink}}

\usepackage{scalerel}
\usepackage{xurl}
\usepackage[colorlinks=true, urlcolor=blue, citecolor=blue,linktoc=page]{hyperref}

\usepackage{tablefootnote}

\newcommand{\ee}{\end{equation}}
\newcommand{\beq}{\begin{equation}}
\newcommand{\eeq}{\end{equation}}
\newcommand{\bes}{\begin{eqnarray}}
\newcommand{\ees}{\end{eqnarray}}
\newcommand{\bqa}{\begin{eqnarray}}
\newcommand{\eqa}{\end{eqnarray}}
\newcommand{\bea}{\begin{eqnarray}}
\newcommand{\eea}{\end{eqnarray}}

\newcommand{\R}{\mathbb{R}}

\newcommand{\Z}{\mathbb{Z}}

\newcommand{\cD}{{\cal D}}

\newcommand{\cF}{{\cal F}}

\newcommand{\cO}{{\cal O}}

\newcommand{\cW}{{\cal W}}

\newcommand{\red}{\mathrm{red}}
\newcommand{\pgf}{\mathrm{pgf}}
\newcommand{\BF}{\mathrm{BF}}

\DeclareMathOperator{\Vol}{Vol}

\def\fg{\mathfrak{g}}

\newcommand{\bbm}{\begin{bmatrix}}
\newcommand{\ebm}{\end{bmatrix}}

\usepackage{bbding}

\newlength\myheight  

\theoremstyle{definition}
\newtheorem{theorem}{Theorem}[section]
\AtEndEnvironment{theorem}{\null\hfill\ensuremath{\scaleto{\diamond}{\myheight}}}

\AtEndEnvironment{definition}{\null\hfill\ensuremath{\scaleto{\diamond}{\myheight}}}

\newtheorem{proposition}[theorem]{Proposition}
\AtEndEnvironment{proposition}{\null\hfill\ensuremath{\scaleto{\diamond}{\myheight}}}

\AtEndEnvironment{example}{\null\hfill\ensuremath{\scaleto{\diamond}{\myheight}}}

\newtheorem{lemma}[theorem]{Lemma}
\AtEndEnvironment{lemma}{\null\hfill\ensuremath{\scaleto{\diamond}{\myheight}}}

\newtheorem{remark}[theorem]{Remark}
\AtEndEnvironment{remark}{\null\hfill\ensuremath{\scaleto{\diamond}{\myheight}}}

\AtEndEnvironment{assumption}{\null\hfill\ensuremath{\scaleto{\diamond}{\myheight}}}

\AtEndEnvironment{corollary}{\null\hfill\ensuremath{\scaleto{\diamond}{\myheight}}}

\title{Regge Dilaton Gravity from Feynman Diagrams}

\author[1]{Florian Girelli\footnote{florian.girelli@uwaterloo.ca}}
\author[1,2]{Christopher Pollack\footnote{cajpolla@uwaterloo.ca}}

\affil[1]{\textit{Department of Applied Mathematics, University of Waterloo, 200 University Avenue West, Waterloo, Ontario, Canada, N2L 3G1}}
\affil[2]{\textit{Perimeter Institute for Theoretical Physics, 31 Caroline St. N., Waterloo, Ontario, Canada, N2L 2Y5}}
\date{\today}

\begin{document}

\maketitle

\begin{abstract}
Baratin and Freidel showed that closed Feynman diagrams in three and four Euclidean dimensions can be expressed as expectation values of observables in a Regge-type topological model \cite{baratin_hidden_2007, baratin_hidden_2007-1}. We extend their analysis to two dimensions with spherical, flat, or hyperbolic background geometry. Specifically, we show that the reduced Feynman amplitude of any closed (planar) scalar graph can be represented as an observable insertion in the partition function of either a discretized BF theory, whose gauge group is the isometry group of the background, or a homogeneous Regge discretization of linear dilaton gravity.

\end{abstract}

\tableofcontents

\section*{Introduction}\addcontentsline{toc}{section}{Introduction}

Feynman amplitudes on homogeneous spacetime backgrounds admit a geometric interpretation that goes beyond their usual role in perturbative quantum field theory. In a series of seminal papers \cite{baratin_hidden_2007, baratin_hidden_2007-1}, building on earlier work by Barrett \cite{barrett_feynman_2006}, Baratin and Freidel showed how Regge-type topological models can be extracted from Feynman diagrams in three- and four-dimensional Euclidean spacetimes.

The key observation is that the propagators depend only on the relative distances between their endpoints. These distances, however, are not all independent, but satisfy geometric constraints inherited from the homogeneous background. Exploiting these constraints allows the Feynman amplitudes to be reformulated in terms of a Regge-type topological model whose fundamental variables are relative lengths.

To make this more precise, consider a scalar field theory on a homogeneous space $X$, and let $\Gamma$ be a finite connected planar graph with vertex set $V_\Gamma$, $|V_\Gamma|=N\geq 2$, edge set $E_\Gamma$, and no external legs.\footnote{Planarity is assumed so that $\Gamma$ can be embedded in the one-skeleton of a triangulation.} Applying the standard Feynman rules, the corresponding closed Feynman amplitude takes the following position-space form,\footnote{In general, coupling constants and graph automorphism factors multiply \eqref{eq:intro-feynman}. Since they play no role in the geometric reconstructions considered here, we suppress them.}
\begin{equation}
  I_\Gamma^X
  =\int_{X^N}\prod_{i=1}^{N}d\mu_X(x_i)
  \prod_{(ij)\in E_\Gamma}G_F\bigl(d_X(x_i,x_j)\bigr),
  \label{eq:intro-feynman}
\end{equation}
where $d\mu_X$ denotes the invariant measure on $X$, $d_X$ its invariant distance, and $G_F$ a regulated Feynman propagator. The Baratin--Freidel construction consists of rewriting such amplitudes in terms of relative lengths, while accounting for the geometric constraints among them.

The relation between the resulting topological model and BF theory was highlighted in three dimensions for the flat case \cite{baratin_hidden_2007},  conjectured in \cite{baratin_hidden_2007, baratin_hidden_2007-1} and subsequently established explicitly in four dimensions \cite{asante_quantum_2020}. More recently, we constructed the two-dimensional analog of the Baratin--Freidel model and showed that, for $\Lambda\leq 0$, it can be obtained from a suitable discretization of BF theory \cite{girelli_regge_2026}. An interesting feature of the two-dimensional setting is that this BF theory also coincides with the first order formulation of linear dilaton gravity \cite{mertens_solvable_2023}.

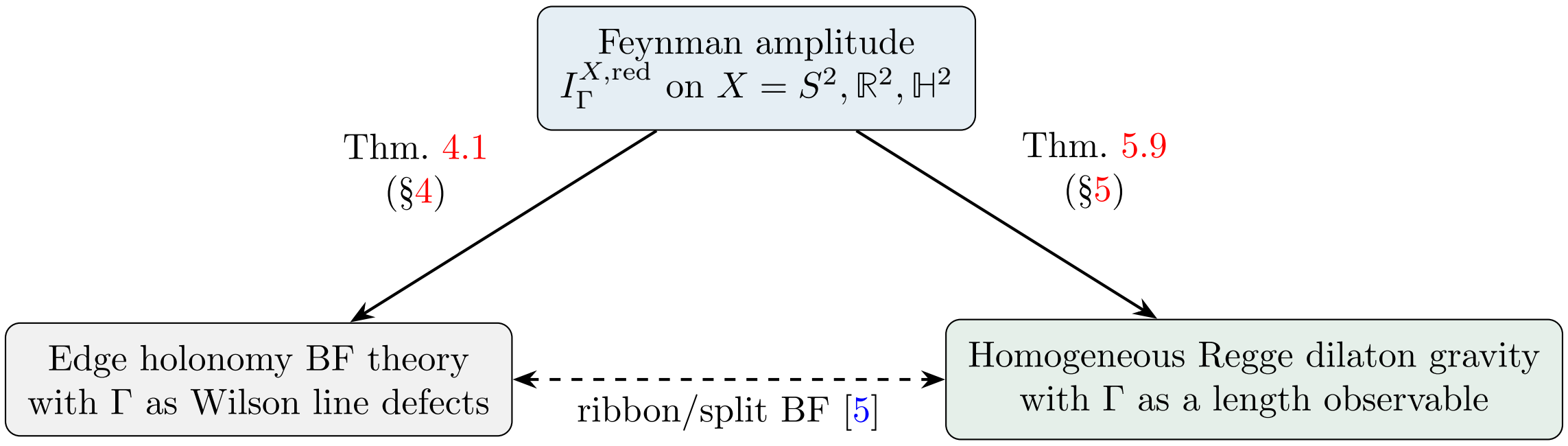
\begin{figure}[h]
\centering
\begin{tikzpicture}[font=\small,>=Stealth]
 \node[draw,rounded corners,fill=jtBlue,align=center,inner sep=6pt] (F) at (0,0) {Feynman amplitude\\ $I^{X,\red}_{\Gamma}$ on $X=S^{2},\R^{2},\mathbb{H}^{2}$};
 \node[draw,rounded corners,fill=jtGrey,align=center,inner sep=6pt] (B) at (-4.8,-3.0) {Edge holonomy BF theory\\ with $\Gamma$ as Wilson line defects};
 \node[draw,rounded corners,fill=jtSage,align=center,inner sep=6pt] (R) at (4.8,-3.0) {Homogeneous Regge dilaton gravity\\ with $\Gamma$ as a length observable};
 \draw[->,thick] (F) -- node[above left,align=center] {Thm.\ \ref{thm:BF-recovery-from-feyn-amp}\\ (\S\ref{sec:bf-reconstruction})} (B);
 \draw[->,thick] (F) -- node[above right,align=center] {Thm.\ \ref{thm:main}\\ (\S\ref{sec:regge-reconstruction})} (R);
 \draw[<->,thick,dashed] (B) -- node[below,align=center] {ribbon/split BF \cite{girelli_regge_2026}} (R);
\end{tikzpicture}
\caption{The three descriptions related in this work. The two solid arrows are established here for all signs of the cosmological constant, the dashed arrow is the BF to Regge correspondence of \cite{girelli_regge_2026}, established there for Euclidean signature, $\Lambda\leq 0$, non-folded triangles and vanishing deficit angles (the zero winding sector of \cite{girelli_regge_2026}, i.e., $w_{v} = 1$ in \eqref{eq:angular-closure}).}
\label{fig:triangle-of-relations}
\end{figure}

The aim of the present work is to complete this picture by deriving the two-dimensional Baratin--Freidel model directly from Feynman diagrams. Together with our previous results \cite{girelli_regge_2026}, this establishes a correspondence between Feynman amplitudes on homogeneous backgrounds, topological BF theory, Regge-type discretizations, and two-dimensional dilaton gravity (see Figure \ref{fig:triangle-of-relations}).

The emergence of a topological theory from Feynman amplitudes in this context is, in fact, quite natural. In Section \ref{sec:gauging-isometries-and-topological-defects}, we first explain how the geometry of a homogeneous spacetime background can be promoted to dynamical variables through a BF theory. Since the BF gauge group is the isometry group of the background, this construction can be understood as a gauging of spacetime isometries.
As recalled in Section \ref{subsec:particles-as-defects}, following \cite{Balachandran:1983oit,freidel_particles_2006}, the phase space of a particle propagating on a homogeneous spacetime can be described in terms of a coadjoint orbit of the spacetime isometry group. This provides a natural group-theoretic description of particle dynamics, with the corresponding orbits labeled by a Casimir of the isometry algebra. Upon quantization, this picture leads to an interpretation of particle propagation in terms of Wilson lines coupled to the topological theory.

The same gauging of spacetime isometries admits a direct realization at the level of Feynman diagrams. In Section \ref{sec:bf-reconstruction}, we show how the homogeneity of the background allows the vertex data of a Feynman graph to be reconstructed in terms of holonomies of the isometry group. This leads to a discretized BF partition function, in which the flatness of the connection is imposed through holonomy constraints \cite{Birmingham:1991ty, fukuma_lattice_1994, livine_2d_2001, oeckl_generalized_2003, oriti_spinfoam_2005, ozer_holonomies_2026}. In this formulation, the original Feynman graph can be interpreted as a network of Wilson lines, with the appropriate propagator decorations.

Finally, in Section \ref{sec:regge-reconstruction}, we return to the original Baratin--Freidel construction and extend it to two dimensions. By exploiting the geometric constraints among relative lengths, we recover directly from the Feynman amplitudes the two-dimensional Regge-type topological invariant. As established in our companion paper \cite{girelli_regge_2026}, for $\Lambda \leq 0$, restricted to consistently oriented (non-folded) triangles and to deficit angles vanishing exactly rather than modulo $2\pi$, the same model arises from a suitable discretization and partial gauge fixing of BF theory, which also provides the first order formulation of linear dilaton gravity. This completes the correspondence between the Feynman, BF, and Regge formulations, and shows how the geometric structure underlying Feynman diagrams can be related to two-dimensional dilaton gravity.

We begin by fixing our conventions.

\section{Setup}

Throughout, we adopt the conventions of \cite{girelli_regge_2026}, which we briefly recall.

We consider a two-dimensional spacetime $(M,g_{\mu\nu})$, with spacetime indices $\mu,\nu=0,1$, locally of constant curvature
\begin{equation}
    \Lambda=\frac{\kappa}{l^2},
    \quad \kappa\in\{-1,0,1\},
    \label{eq:cosmo-constant}
\end{equation}
where $\kappa$ specifies the curvature sign and $l=1/\sqrt{|\Lambda|}$ for $\Lambda\neq0$. In the flat case, $l$ is arbitrary. In the first order formulation, the geometry is encoded in a zweibein $e^a$ and a spin connection $\omega$, with internal indices $a,b=0,1$ and
\begin{equation}
    g_{\mu\nu}=\eta_{ab}e^a_\mu e^b_\nu,
    \quad
    \eta_{ab}=\mathrm{diag}(\sigma,1),
    \quad \sigma=\pm1.
\end{equation}
The Euclidean and Lorentzian signatures correspond respectively to $\sigma=1$ and $\sigma=-1$. We use the conventions
\begin{equation}
    \epsilon_{01}=1,
    \quad
    \epsilon^{01}=\sigma,
    \quad
    \epsilon_{ab}\epsilon^{bc}=-\sigma\delta_a^c.
\end{equation}

Each pair $(\sigma,\kappa)$ determines a constant curvature model space $X$, as summarized in Table \ref{table:2d-spacetimes}. More generally, $M$ is locally modeled on $X$, with coordinate charts related by local isometries of $X$ \cite{goldman_locally_2010}.

\begin{table}[h]
    \centering
    \begingroup
    \setlength{\tabcolsep}{14pt}
    \renewcommand{\arraystretch}{1.2}
    \begin{tabular}{@{}lccc@{}}
        \toprule
        Signature/curvature & $\kappa=1$ & $\kappa=0$ & $\kappa=-1$ \\
        \midrule
        Euclidean ($\sigma=1$)
            & $S^2$ & $\mathbb{R}^2$ & $\mathbb{H}^2$ \\
        Lorentzian ($\sigma=-1$)
            & $\mathrm{dS}_2$ & $\mathbb{R}^{1,1}$ & $\mathrm{AdS}_2$ \\
        \bottomrule
    \end{tabular}
    \endgroup
    \caption{Two-dimensional constant curvature model spaces $X$, classified by signature $\sigma$ and curvature sign $\kappa$.}
    \label{table:2d-spacetimes}
\end{table}

Denoting by $G_{\sigma,\kappa}$ the isometry group of $X$, and by $H_\sigma\subset G_{\sigma,\kappa}$ the stabilizer of a reference point $x_0\in X$, the model space admits the homogeneous space description
\begin{equation}
    X\simeq G_{\sigma,\kappa}/H_\sigma,
    \quad
    H_\sigma=\{g\in G_{\sigma,\kappa}\mid g\cdot x_0=x_0\},
    \label{eq:X-as-quotient}
\end{equation}
where $H_1=SO(2)$ and $H_{-1}=SO(1,1)$. Explicitly,
\begin{subequations}
\begin{align}
    S^2 &\cong SO(3)/SO(2),
    &
    \mathbb{R}^2 &\cong ISO(2)/SO(2),
    &
    \mathbb{H}^2 &\cong SO(2,1)/SO(2),
    \\
    \mathrm{dS}_2 &\cong SO(1,2)/SO(1,1),
    &
    \mathbb{R}^{1,1} &\cong ISO(1,1)/SO(1,1),
    &
    \mathrm{AdS}_2 &\cong SO(2,1)/SO(1,1).
\end{align}
\end{subequations}
The invariant metric on $X$ determines an invariant measure $d\mu_X$ and a corresponding invariant distance (or interval) $d_X(\bullet,\bullet)$, which enter the Feynman amplitude \eqref{eq:intro-feynman}.\footnote{In Lorentzian signature, the invariant interval is not positive definite and does not define a global metric-space distance.}

\section{Dilaton gravity from the scalar field action}\label{sec:gauging-isometries-and-topological-defects}

Consider a scalar field $\varphi$ on a fixed homogeneous background with action $S_{\mathrm m}[\varphi;e]$. Its action depends on the background only through the zweibein $e^{a}$, and the isometries of the background are global symmetries. In this section we recall how promoting the background to a dynamical field while enforcing homogeneity of the geometry through Lagrange multipliers produces two-dimensional BF theory with gauge group $G_{\sigma,\kappa}$, wherein the isometries of the background are thereby turned into gauge symmetries.

The key point is that the geometric data of the homogeneously curved manifold $(M,g_{\mu\nu})$ is encoded in a zweibein $e^{a}$ and spin connection $\omega$ satisfying the relations,
\begin{equation}
g_{\mu\nu} = \eta_{ab}e^{a}_{\mu}e^{b}_{\nu},\quad d\omega+\sigma\frac{\kappa}{2l^2}\epsilon_{ab}e^a\wedge e^b = 0,\quad de^{a}+\omega\wedge e^{b}\epsilon_{b}{}^{a} =0 .   \label{eq:homo-M-e-omega-relations}
\end{equation}
These relations can be naturally encoded in a Cartan formulation of the geometry \cite{wise_symmetric_2009}. We package the $(\omega, e^{a})$ data together into the Cartan connection valued in $\mathfrak{g}_{\sigma,\kappa}$, the three-dimensional Lie algebra of $G_{\sigma,\kappa}$, with  rotation (or boost) generator $J$ and the two translation generators $P_{a}$. The Lie brackets read
\begin{equation}
[J,P_a]=\epsilon_a{}^bP_b,
\quad
[P_a,P_b]=\sigma\frac{\kappa}{l^2}\epsilon_{ab}J = \sigma\Lambda\epsilon_{ab}J.
\label{eq:isometry_algebra}
\end{equation}
Explicitly, we have
\begin{subequations}
\begin{align}
\fg_{1,1} \cong \mathfrak{so}(3) ,\quad \fg_{1,0} &\cong \mathfrak{iso}(2),\quad \fg_{1,-1} \cong \mathfrak{so}(2,1),\label{eq:euclidean-cartan-lie-alg}\\
    \fg_{-1,1} \cong \mathfrak{so}(1,2),\quad \fg_{-1,0} &\cong \mathfrak{iso}(1,1),\quad \fg_{-1,-1} \cong \mathfrak{so}(2,1).\label{eq:lorentzian-cartan-lie-alg}
\end{align}
\end{subequations}
The Cartan connection is therefore 
\begin{equation}
    A:= \omega J + e^{a}P_{a}\in \Omega^{1}(M,\mathfrak{g}_{\sigma,\kappa}).\label{eq:A-cartan}
\end{equation}
Its associated curvature 2-form $\cF(A) = dA+\frac{1}{2}\comm{A}{A}$  decomposes as
\begin{equation}
\mathcal{F}(A)=FJ+T^aP_a ,\label{eq:A-curvature-decomp}
\end{equation}
with
\begin{subequations}{\label{eq:2d_torsion_curvature}}
\begin{align}
F &= d\omega+\sigma\frac{\kappa}{2l^2}\epsilon_{ab}e^a\wedge e^b\\
T^a &= de^a+\omega\wedge e^b\epsilon_{b}{}^{a}.\label{eq:cartan-torsion}
\end{align}
\end{subequations}
Comparing to \eqref{eq:homo-M-e-omega-relations}, we see the geometry of $M$ is characterized precisely by the vanishing of $\cF(A)$, or equivalently the vanishing of $F$ and $T^{a}$,
\begin{equation}
\cF(A) = 0 \iff F = 0 \text{ and } T^{a}=0.\label{eq:homogeneous_constraints}
\end{equation}
Equivalently, the geometry of $M$ is torsionless and has constant scalar curvature
\begin{equation}
R(e)=2\frac{\kappa}{l^2}.
\label{eq:constant_scalar_curvature}
\end{equation}
Indeed, imposing $T^{a} =0$ gives $\omega\mapsto \omega(e)$, such that imposing $F = 0$ and using the standard relations
\begin{equation}
d\omega(e)=-\frac{\sigma}{2}R(e)\,\epsilon,
\quad
\epsilon=\frac12\epsilon_{ab}e^a\wedge e^b ,\label{eq:standard-relations}
\end{equation}
implies \eqref{eq:constant_scalar_curvature}.

Gauging the isometries means promoting the background fields $(e^a,\omega)$ to be dynamical variables, all while imposing \eqref{eq:homogeneous_constraints} by Lagrange multipliers, which we will denote $B$. 

\medskip 

Therefore alongside $\fg_{\sigma,\kappa}$ we need its vector space dual $\fg^{*}$, regarded as a real three-dimensional \textit{Abelian} Lie algebra. Its generators $\tilde{J}$ and $\tilde{P}^{a}$ are dual to $J$ and $P_{a}$ with respect to the canonical pairing $\langle \bullet,\bullet\rangle$ between $\fg_{\sigma,\kappa}$ and $\fg^{*}$ \cite{snobl_classification_2002},
\begin{equation}
    \langle J,\tilde{J}\rangle = 1,\quad \langle P_{a},\tilde{P}^{b}\rangle = \delta_{a}^{b},\quad \langle J,\tilde{P}^{a}\rangle = \langle P_{a},\tilde{J}\rangle = 0,\label{eq:g-g*-Cartan-pairing}
\end{equation}
and $\fg^{*}$ carries the (dressing) action $\lhd$ of $\fg_{\sigma,\kappa}$ induced by the coadjoint action,\footnote{Note there is no ``back-action'' of $\fg^{*}$ on $\fg_{\sigma,\kappa}$. It may also be useful to recall that the dressing action has the anti-symmetric property $\alpha\lhd \beta = -\beta\rhd \alpha$ for all $\beta\in \fg_{\sigma,\kappa}$ and $\alpha\in \fg^{*}$.}
\begin{equation}
\tilde{J}\lhd J = 0,\quad \tilde{J}\lhd P_{a} = \sigma\frac{\kappa}{l^{2}}\epsilon_{ab}\tilde{P}^{b},\quad \tilde{P}^{a}\lhd J = \epsilon_{b}{}^{a}\tilde{P}^{b},\quad \tilde{P}^{a}\lhd P_{b} = -\epsilon_{b}{}^{a}\tilde{J}.\label{eq:Cartan-Dressing-Lie-Alg}
\end{equation}
Explicitly, the $B$-field 
\begin{equation}
B=\Phi \tilde{J}+X_a\tilde{P}^a \in \Omega^{0}(M,\fg^{*})
\label{eq:2d_B_field}
\end{equation}
plays the role of the desired Lagrange multiplier, such that along with the invariant pairing the two-dimensional BF action
\begin{equation}
S_{\mathrm{BF}}[A,B]
=
\int_M\langle B,\cF(A)\rangle
=
\int_M
X_aT^a+\Phi F
\label{eq:2d_bf_action}
\end{equation}
admits exactly the dynamics \eqref{eq:homogeneous_constraints} when varying with respect to $B$ and has $G_{\sigma,\kappa}$ as its gauge group.

\medskip 

Consider now a minimally coupled scalar matter field $\varphi$ propagating on the fixed homogeneous background $X$. For a generic potential $V(\varphi)$, its action is given by\footnote{This is nothing more than the standard action
\begin{equation*}
    S_{m}[\varphi;g] = -\frac{1}{2}\int_{M}d^{2}x\sqrt{|g|}g^{\mu\nu}\partial_{\mu}\varphi \partial_{\nu}\varphi-\int_{M}d^{2}x\sqrt{|g|}V(\varphi)
\end{equation*}
for a minimally coupled scalar field propagating on a fixed spacetime background, written in first order metric variables.}
\begin{equation}
S_{\mathrm m}[\varphi;e]
=
-\frac12
\int_M
d^2x\,|e|\,
\eta^{ab}E_a{}^\mu E_b{}^\nu
\partial_\mu\varphi\,\partial_\nu\varphi
-
\int_M
d^2x\,|e|\,V(\varphi),
\label{eq:scalar_background_action}
\end{equation}
where $e=\det(e^a{}_\mu)$, and $E_a{}^\mu$ is the inverse frame. Since $\varphi$ is a Lorentz scalar, the spin connection $\omega$ enters only through the homogeneous geometry conditions \eqref{eq:homogeneous_constraints}.

The matter-coupled action with dynamical $(e^{a},\omega)$ is then simply given by
\begin{equation}
S[e,\omega,X,\Phi;\varphi]
=
S_{\mathrm{BF}}[A,B]
+
S_{\mathrm m}[\varphi;e].
\label{eq:matter_bf_action}
\end{equation}
We henceforth adopt the notation ``$\approx$'' to mean going on-shell. The full set of equations of motion for the matter-coupled action \eqref{eq:matter_bf_action} is given by,
\begin{subequations}{\label{eq:BF-matter-coupled-EOMs}}
\begin{align}
T^a&\approx 0,
\\
F&\approx 0,
\\
d B+A\rhd B&\approx j_{\mathrm m}, \label{DAB=J}
\\
\frac{1}{e}\partial_\mu
\left(
e\,g^{\mu\nu}\partial_\nu\varphi
\right)
-
V'(\varphi)&\approx 0.
\end{align}
\end{subequations}
Equation \eqref{DAB=J} is a manifestation of the remark just after \eqref{eq:scalar_background_action}, i.e., for a minimally coupled scalar there is no independent spin current, hence the Lorentz component of $j_{\mathrm m}$ vanishes while the translational component is the energy-momentum current. Thus matter does not deform the flatness constraints themselves, but instead sources the Lagrange multiplier equation $d_AB\approx j_{\mathrm m}$.

The familiar two-dimensional dilaton gravity action in the constant curvature, linear potential sector is recovered from \eqref{eq:2d_bf_action} simply by integrating out $\omega$. Indeed, varying the BF action with respect to $X_a$ gives $T^{a}\approx 0$, which for an invertible zweibein $e^{a}$, fixes $\omega$ to be the Levi-Civita spin connection $\omega(e)$. Substituting this back into \eqref{eq:2d_bf_action}, and applying the standard relations \eqref{eq:standard-relations}, one obtains, up to boundary terms and an overall normalization,
\begin{equation}
S_{\mathrm{BF}}
\mapsto 
\frac12
\int_M
d^2x\,\sqrt{|g|}\,
\Phi
\left(
R(e)-\frac{2\kappa}{l^2}
\right).
\label{eq:jt_from_bf}
\end{equation}
This is the linear dilaton gravity (LDG) action with homogeneous curvature sign $\kappa$ \cite{mertens_solvable_2023,ikeda_general_1993,ikeda_two-dimensional_1994}. The scalar $\Phi$ is called the dilaton field. Variation with respect to $\Phi$ imposes
\begin{equation}
R(e)-\frac{2\kappa}{l^2}\approx 0.
\end{equation}
The remaining first order equations are
\begin{subequations}
\begin{align}
d\Phi+\epsilon_{b}{}^{a}X_a e^b&\approx 0,
\label{eq:eom_omega}
\\
d X_a
-\omega \epsilon_{a}{}^{b}X_{b}+
\sigma\frac{\kappa}{l^2}\Phi\,\epsilon_{ab}e^b&\approx 0.
\label{eq:eom_frame}
\end{align}
\end{subequations}
The first equation identifies $X_a$ with the frame derivative of the dilaton,
\begin{equation}
X_a=\sigma\epsilon_a{}^bE_b{}^\mu\partial_\mu\Phi.
\label{eq:Xa_dilaton_derivative}
\end{equation}
Substituting this into \eqref{eq:eom_frame} and converting to second order metric variables gives the metric dilaton equation
\begin{equation}
\nabla_\mu\nabla_\nu\Phi
-
g_{\mu\nu}\nabla^2\Phi
-
\frac{\kappa}{l^2}g_{\mu\nu}\Phi
\approx 0.
\label{eq:metric_dilaton_eom}
\end{equation}
Thus the BF equations reproduce the LDG equations in the constant curvature sector.

In summary, the $\tilde{J}$ component $\Phi$ of the $B$ field is the dilaton, and the flatness constraints of $G_{\sigma,\kappa}$ BF theory are the equations of motion of constant curvature LDG. The discrete counterpart of this statement is the identification, in \eqref{eq:Regge-state-sum} below, of the integer labels attached to the vertices of the Regge model with a discrete dilaton: the Regge action which emerges from the Feynman amplitude is $\sum_{v}s_{v}\delta_{v}^{\epsilon}$, with $\delta_{v}^{\epsilon}$ the signed deficit angle \eqref{eq:oriented-deficit} at $v$, i.e., an oriented version of the discretization of \eqref{eq:jt_from_bf} studied in \cite{girelli_regge_2026}.

\section{Dilaton gravity from propagators}\label{subsec:particles-as-defects}

We now discuss how from the definition of the propagator for a scalar field we can also extract some dilaton gravity type information.

Let $\varphi$ be a scalar field of mass $m$ on a two-dimensional background with zweibein $e^a$. Let us recall how its propagator can then be related to a Wilson line. The Feynman propagator is the inverse of the Klein--Gordon operator,\footnote{Equations \eqref{eq:scalar_propagator_operator}--\eqref{eq:scalar_worldline_second_order} are written for $\sigma=1$. For $\sigma=-1$, \eqref{eq:scalar_propagator_operator} is understood with $m^2\to m^2-i\varepsilon$, which selects the Feynman propagator, and \eqref{eq:schwinger_proper_time}--\eqref{eq:scalar_worldline_second_order} follow by rotating $T\to iT$ and $\tau\to i\tau$. The first order form \eqref{eq:scalar_worldline_first_order} holds in both signatures.}
\begin{equation}
G_m(x_f,x_i\mid e)
=
\left\langle x_f\left|
\left(-\nabla_g^2+m^2\right)^{-1}
\right|x_i\right\rangle .
\label{eq:scalar_propagator_operator}
\end{equation}
Using the Schwinger proper-time representation, one may write \cite{polyakov_quantum_1981, bastianelli_path_2005}
\begin{equation}
G_m(x_f,x_i\mid e)
=
\int_0^\infty dT\,
\left\langle x_f\left|
e^{-T(-\nabla_g^2+m^2)}
\right|x_i\right\rangle .
\label{eq:schwinger_proper_time}
\end{equation}
The heat kernel admits the worldline representation
\begin{equation}
G_m(x_f,x_i\mid e)
=
\int_0^\infty dT
\int_{z(0)=x_i}^{z(T)=x_f}
\mathcal Dz\,
\exp\left[
-\int_0^T d\tau
\left(
\frac{1}{4}\,
g_{\mu\nu}(z)\dot z^\mu\dot z^\nu
+
m^2
\right)
\right].
\label{eq:scalar_worldline_second_order}
\end{equation}
Equivalently, in first order form,
\begin{equation}
G_m(x_f,x_i\mid e)
=
\int
\mathcal Dz\,\mathcal Dp\,\mathcal DN\,
\exp\left\{
i\int d\tau
\left[
p_a e^a_\mu(z)\dot z^\mu
-
\frac{N}{2}
\left(
\eta^{ab}p_ap_b+m^2
\right)
\right]
\right\},
\label{eq:scalar_worldline_first_order}
\end{equation}
where $N(\tau)>0$ is the worldline einbein imposing the mass-shell constraint
\begin{equation}
\eta^{ab}p_ap_b+m^{2}=0 .\label{eq:mass_shell_constraint}
\end{equation}
The action in the exponential is the relativistic particle action in a homogeneous spacetime. Such an action can be formulated in terms of coadjoint orbits of $G_{\sigma,\kappa}$ \cite{Balachandran:1983oit, freidel_particles_2006}. Let us recall the construction. The particle worldline is
\begin{equation}
    W:\tau\mapsto z^\mu(\tau),
\end{equation}
and its intrinsic charge is a fixed element of $\fg^{*}$,
\begin{equation}
K=m\tilde{P}^0+s\tilde{J} .
\label{eq:2d_rest_charge}
\end{equation}
The coadjoint orbit through $K$ is labeled by the value of the invariant $\eta^{ab}p_{a}p_{b}+\sigma\frac{\kappa}{l^{2}}s^{2}$ of $p_{a}\tilde{P}^{a}+s\tilde{J}\in \fg^{*}$, i.e., by a single combination of $m$ and $s$.\footnote{The identification of \eqref{eq:scalar_worldline_first_order} with the orbit through $K$ holds in Lorentzian signature with $s=0$, where both give $\eta^{ab}p_ap_b=-m^2$. In Euclidean signature \eqref{eq:mass_shell_constraint} has no real solutions, and \eqref{eq:scalar_worldline_first_order} integrates over all momenta, i.e., over all orbits, rather than over a single one.} In two dimensions the spin is a Lorentz scalar because the Lorentz algebra is one-dimensional. 

The dynamical variable on the worldline $W$ is a group element 
\begin{equation}
h(\tau)\in H_{\sigma}\subset G_{\sigma,\kappa},
\end{equation}
i.e., a rotation (or boost) of the rest frame of the particle. The dressed charge is
\begin{equation}
\mathcal J(\tau):=h(\tau)\rhd K=p_a(\tau)\tilde{P}^a+s\tilde{J},\quad \eta^{ab}p_ap_b=\sigma m^{2},
\label{eq:2d_dressed_charge}
\end{equation}
where by \eqref{eq:Cartan-Dressing-Lie-Alg}, $H_{\sigma}$ leaves $\tilde{J}$ invariant and transforms the $\tilde{P}^{a}$ among themselves. Hence the spin $s$ is constant and $p_{a}(\tau)$ sweeps the $H_{\sigma}$ orbit of $m\tilde{P}^{0}$, so that $\mathcal{J}(\tau)$ explores only a one-dimensional submanifold of the coadjoint orbit through $K$. 

The particle action is the coadjoint orbit action minimally coupled to the
connection $A$:
\begin{equation}
S_W[z,h;A]
= \int_W \langle K,A^h_\tau\rangle d\tau, \quad A^h_\tau
:=
h^{-1}A_\mu(z)h\,\dot z^\mu+h^{-1}\dot h ,
\label{eq:2d_particle_action}
\end{equation}
where the overdot denotes differentiation with respect to $\tau$. Equivalently using invariance of the pairing,
\begin{equation}
S_W[z,h;A]
=
\int d\tau
\Big(
\langle\mathcal J,A_\mu(z)\rangle\dot z^\mu
+
\langle K,h^{-1}\dot h\rangle
\Big).
\label{eq:2d_particle_action_J}
\end{equation}
Plugging in the decomposition \eqref{eq:A-cartan} of $A$, the above becomes
\begin{equation}
S_W
=
\int_W d\tau
\Big(
p_a e^a_\mu(z)\dot z^\mu
+
s\,\omega_\mu(z)\dot z^\mu
+
\langle K,h^{-1}\dot h\rangle
\Big).
\label{eq:2d_particle_component_action}
\end{equation}
Thus $p_a$ couples to the zweibein and $s$ couples to the spin
connection.

The total action of a point particle coupled to two-dimensional BF theory (i.e., dilaton gravity in the first order formalism) is then given by
\begin{equation}
S[A,B;z,h]
=
S_{\mathrm{BF}}[A,B]+S_W[z,h;A].
\label{eq:2d_total_action}
\end{equation}
Variation with respect to $B$ gives $\cF(A)\approx 0$, still encoding the background geometry structure. Variation with respect to $A$ gives the sourced BF equation
\begin{equation}
d_AB\approx \mathcal J\,\delta_W,
\label{eq:2d_sourced_BF}
\end{equation}
where $\delta_W$ is the one-form distribution Poincar\'e dual to the worldline
\begin{equation}
\int_M\delta_W\wedge\alpha
=
\int_W
\alpha_\mu(z(\tau))\dot z^\mu(\tau)d\tau .
\label{eq:2d_deltaP}
\end{equation}
In components the sourced BF equation \eqref{eq:2d_sourced_BF} gives,
\begin{subequations}
\begin{align}
d\Phi+\epsilon_{b}{}^{a}X_{a}e^{b}
&\approx 
s\,\delta_W,
\label{eq:2d_spin_source}
\\
d X_a-\omega\epsilon_{a}{}^{b}X_{b}+\sigma\frac{\kappa}{l^2}\epsilon_{ab}e^b\Phi
&\approx
p_a\,\delta_W.
\label{eq:2d_momentum_source}
\end{align}
\end{subequations}
We see that the scalar spin sources the Lorentz equation, while the momentum sources the
translational equation. This is consistent with the equations of motion \eqref{eq:BF-matter-coupled-EOMs} of the matter-coupled-BF we discussed in \S\ref{sec:gauging-isometries-and-topological-defects}. 

To write the remaining equations of motion we introduce the covariant derivative of the dressed charge along the worldline,
\begin{equation}
D_\tau\mathcal J:=\frac{d\mathcal J}{d\tau}-\mathcal{J}\lhd A_{\tau}
=\Big(\nabla_\tau p_a+s\sigma\frac{\kappa}{l^2}\,\epsilon_{ab}e^b_\tau\Big)\tilde{P}^{a}
+\epsilon_{b}{}^{a}p_{a}e^b_\tau\,\tilde{J},
\quad A_\tau:=A_\mu(z)\dot z^\mu,
\label{eq:2d_charge_transport}
\end{equation}
where $e^a_\tau:=e^a_\mu\dot z^\mu$ and 
\begin{equation}
    \nabla_\tau p_a=\dot p_a-\omega_\tau\epsilon_a{}^b p_b.
\end{equation}
Since $h$ takes values in $H_{\sigma}$, its variation gives only the $\tilde{J}$ component of $D_{\tau}\mathcal{J}\approx0$,
\begin{equation}
\epsilon_{b}{}^{a}p_{a}e^b_\tau\approx 0,
\label{eq:2d_spin_parallel}
\end{equation}
i.e., $p_{a}\propto \eta_{ab}e^{b}_{\tau}$ (the momentum is parallel to the velocity).

Finally, varying the embedding $z^\mu(\tau)$ gives
\begin{equation}
\left\langle D_\tau\mathcal J,A_\mu\right\rangle
-
\left\langle\mathcal J,\cF_{\mu\nu}(A)\right\rangle\dot z^\nu
\approx 0.
\label{eq:2d_embedding_variation}
\end{equation}
Using \eqref{eq:2d_spin_parallel} and the decomposition \eqref{eq:A-curvature-decomp} in \eqref{eq:2d_embedding_variation}, where the terms proportional to $\kappa$ cancel, one obtains the two-dimensional Mathisson--Papapetrou equation with torsion\footnote{Here $R_{\mu\nu}:=\partial_{\mu}\omega_{\nu}-\partial_{\nu}\omega_{\mu}$ are the spacetime components of the curvature two-form $d\omega$ of the spin connection and $\varepsilon_{\mu\nu}:=\epsilon_{ab}e^{a}_{\mu}e^{b}_{\nu}$ are those of the volume form $\epsilon$ of \eqref{eq:standard-relations}.}
\begin{equation}
(\nabla_\tau p_a)e^a_\mu
\approx 
s\,R_{\mu\nu}\dot z^\nu
+
p_aT^a_{\mu\nu}\dot z^\nu .
\label{eq:2d_MP_torsion}
\end{equation}
In the torsionless sector this reduces to
\begin{equation}
\nabla_\tau p_\mu
\approx 
s\,R_{\mu\nu}\dot z^\nu,
\quad
p_\mu:=p_a e^a_\mu.
\label{eq:2d_MP_torsionless}
\end{equation}
Since in two dimensions $R_{\mu\nu}\propto\varepsilon_{\mu\nu}$, the spin-curvature force is necessarily transverse to the worldline. On-shell of $\cF(A)\approx0$, where $R_{\mu\nu}=-\sigma\frac{\kappa}{l^{2}}\varepsilon_{\mu\nu}$, \eqref{eq:2d_MP_torsionless} is equivalent to the vanishing of the $\tilde{P}^{a}$ components of \eqref{eq:2d_charge_transport}. Hence on the homogeneous background the dressed charge is parallel transported along the worldline, $D_{\tau}\mathcal{J}\approx0$.  For $s=0$, with the momentum parallel to the velocity by \eqref{eq:2d_spin_parallel}, one recovers the geodesic equation
\begin{equation}
\nabla_\tau p_\mu\approx 0.
\end{equation}
In summary, a point particle coupled to two-dimensional BF theory sources the $B$ field equations through its dressed charge, and for $s=0$ its worldline is a geodesic of the homogeneous geometry. 

For a fixed path $z(\tau)$, define the translational Wilson line factor 
\begin{equation}
    \mathcal{W}_{p}[e;z]:=\mathrm{exp}\Big(i\int d\tau p_{a}(\tau)e_{\mu}^{a}(z(\tau))\dot{z}^{\mu}\Big).
\end{equation}
The scalar propagator may therefore be written schematically as a sum over  Wilson line factors weighted by the mass-shell contribution:
\begin{equation}
{
G_m(x_f,x_i\mid e)
=
\int_{z_i=x_i}^{z_f=x_f}
\mathcal Dz\,\mathcal Dp\,\mathcal DN\,
\cW_p[e;z]\,
\exp\left[
-\frac{i}{2}
\int d\tau\,N
\left(
\eta^{ab}p_ap_b+m^2
\right)
\right]
}.
\label{eq:scalar_propagator_as_wilson_sum}
\end{equation}
In this expression the Wilson factor supplies the coupling to the BF connection, while the remaining exponential imposes the standard scalar particle dynamics.

\section{Discretized BF theory from Feynman diagrams}
\label{sec:bf-reconstruction}

We now show that the reduced Feynman amplitude can be rewritten as a partially gauge fixed expectation value in two-dimensional BF theory.

Let $\Delta$ be a triangulation of the disk $D$, containing $\Gamma$ in its one-skeleton. We denote by $V(\Delta), E(\Delta), T(\Delta)$ the vertex, edge and triangular simplex sets of $\Delta$ respectively. To every oriented edge $e=(ij)$ associate a holonomy
\begin{equation}
    g_{ij}\in G_{\sigma,\kappa},
    \quad
    g_{ji}=g_{ij}^{-1},
\end{equation}
and, for every $\tau\in T(\Delta)$, define the (ordered) triangular holonomy
\begin{equation}
    \mathcal H_\tau(g)
    =
    \prod_{e\subset\partial\tau}g_e .
\end{equation}
The discretized two-dimensional BF partition function is \cite{Birmingham:1991ty, fukuma_lattice_1994, livine_2d_2001, oeckl_generalized_2003, oriti_spinfoam_2005,  ozer_holonomies_2026}
\begin{equation}
    Z_{\rm BF}^{\Delta,D}
    =
    \int\prod_{e\in E(\Delta)}dg_e\,
    \prod_{\tau\in T(\Delta)}
    \delta_{G_{\sigma,\kappa}}
    \bigl(\mathcal H_\tau(g)\bigr).
\end{equation}
It is well known that this partition function is invariant under Pachner moves (up to some volume terms). 

The Feynman graph will define the insertion
\begin{equation}
    \mathcal O_\Gamma^X[g]
    :=
    \prod_{(ij)\in E_\Gamma}
    G_F\!\left(
    d_X(x_0,g_{ij}x_0)
    \right).
\end{equation}

\begin{theorem}[BF recovery from Feynman diagrams]\label{thm:BF-recovery-from-feyn-amp}
The reduced Feynman amplitude
\begin{equation}
I^{X,\mathrm{red}}_\Gamma:=I_{\Gamma}^{X}/\mathrm{Vol}(G_{\sigma,\kappa})\label{eq:red-intro-feynman}
\end{equation}
can be written as 
\begin{equation}
    I_\Gamma^{X,\mathrm{red}}
    =
    \langle\mathcal O_\Gamma^X\rangle_{\Delta,\mathrm{BF}}^{\mathrm{pgf}},
\end{equation}
where
\begin{align}
    \langle\mathcal O_\Gamma^X\rangle_{\Delta,\mathrm{BF}}^{\mathrm{pgf}}
    =
    \frac{1}{
    \Vol(G_{\sigma,\kappa})
    \Vol(H_\sigma)^N}
    &\int
    \prod_{v\in V(\Delta)}d\gamma_v
    \prod_{(ij)\in E(\Delta)}dg_{ij}
    \nonumber\\
    &\times
    \delta_{\rm pgf}[\gamma]
    \prod_{(ij)\in E(\Delta)}
    \delta_{G_{\sigma,\kappa}}
    \!\left(\gamma_i g_{ij}\gamma_j^{-1}\right)
    \mathcal O_\Gamma^X[g],
\end{align}
with (partial) gauge fixing factor
\begin{equation}
    \delta_{\rm pgf}[\gamma]
    =
    \prod_{u\in V_{\rm aux}}
    \delta_{G_{\sigma,\kappa}}
    \!\left(\gamma_u(\gamma_u^0)^{-1}\right),
    \quad
    V_{\rm aux}
    :=
    V(\Delta)\setminus V_\Gamma,
\end{equation}
where $\gamma^{0}_{u}\in G_{\sigma,\kappa}$ are fixed reference elements.
\end{theorem}

\begin{proof}
Since
\begin{equation}
    X\simeq G_{\sigma,\kappa}/H_\sigma,
\end{equation}
we write
\begin{equation}
    x_i=\gamma_i x_0,
    \quad
    \gamma_i\in G_{\sigma,\kappa}.
\end{equation}
Using the quotient measure and invariance of the distance,
\begin{equation}
    \int_Xd\mu_X(x_i)
    =
    \frac{1}{\Vol(H_\sigma)}
    \int_{G_{\sigma,\kappa}}d\gamma_i,
    \quad
    d_X(x_i,x_j)
    =
    d_X(x_0,\gamma_i^{-1}\gamma_jx_0),\label{eq:haar-normalization}
\end{equation}
the reduced Feynman amplitude \eqref{eq:red-intro-feynman} becomes
\begin{equation}
    I_\Gamma^{X,\mathrm{red}}
    =
    \frac{1}{
    \Vol(G_{\sigma,\kappa})\Vol(H_\sigma)^N}
    \int\prod_{i\in V_\Gamma}d\gamma_i\,
    \prod_{(ij)\in E_\Gamma}
    G_F\!\left(
    d_X(x_0,\gamma_i^{-1}\gamma_jx_0)
    \right).
\end{equation}
The vertex integration may be extended to all vertices of $\Delta$ by inserting the partial gauge fixing factor $\delta_{\rm pgf}[\gamma]$. Introducing independent edge variables through
\begin{equation}
    1=
    \int dg_{ij}\,
    \delta_{G_{\sigma,\kappa}}
    \!\left(\gamma_i g_{ij}\gamma_j^{-1}\right),
\end{equation}
gives
\begin{align}
    I_\Gamma^{X,\mathrm{red}}
    =
    \frac{1}{
    \Vol(G_{\sigma,\kappa})\Vol(H_\sigma)^N}
    &\int
    \prod_{v\in V(\Delta)}d\gamma_{v}
    \prod_{(ij)\in E(\Delta)}dg_{ij}\,
    \delta_{\rm pgf}[\gamma]
    \prod_{(ij)\in E(\Delta)}
    \delta_{G_{\sigma,\kappa}}
    \!\left(\gamma_i g_{ij}\gamma_j^{-1}\right)
    \mathcal O_\Gamma^X[g].
\end{align}
It remains to identify the edge constraints with the BF flatness constraints. Since $\Delta$ is simply connected,
\begin{equation}
    \mathcal H_\tau(g)=1
    \quad\forall\tau
    \quad\iff\quad
    g_{ij}=\gamma_i^{-1}\gamma_j
\end{equation}
for some vertex variables $\gamma_i$, unique up to a common left multiplication
\begin{equation}
    \gamma_i\longmapsto q\gamma_i,
    \quad q\in G_{\sigma,\kappa}.
\end{equation}
Indeed, flatness makes the ordered product of edge holonomies between two vertices path-independent. Choosing a reference vertex $v_0$ and defining $\gamma_i$ as the parallel transport from $v_0$ to $v_i$ then gives directly
\begin{equation}
    g_{ij}=\gamma_i^{-1}\gamma_j.
\end{equation}
The converse follows immediately by telescoping around each triangle.

The normalization can be seen by choosing a spanning tree of the one-skeleton of $\Delta$. The constraints associated with its $|V|-1$ edges determine all vertex variables in terms of one root variable $\gamma_{v_0}$. The remaining
\begin{equation}
    |E|-|V|+1=|T|
\end{equation}
constraints are equivalent to the face-flatness constraints, where we used Euler's relation for a triangulated disk. The remaining integration over $\gamma_{v_0}$ produces a $\Vol(G_{\sigma,\kappa})$ factor. Hence
\begin{equation}
    \prod_{\tau\in T(\Delta)}
    \delta_{G_{\sigma,\kappa}}
    \!\left(\mathcal H_\tau(g)\right)
    =
    \frac{1}{\Vol(G_{\sigma,\kappa})}
    \int\prod_{v\in V(\Delta)}d\gamma_v\,
    \prod_{(ij)\in E(\Delta)}
    \delta_{G_{\sigma,\kappa}}
    \!\left(\gamma_i g_{ij}\gamma_j^{-1}\right).
\end{equation}
This identifies the previous expression precisely with the partially gauge fixed BF expectation value of $\mathcal O_\Gamma^X[g]$.
\end{proof}

This formulation also clarifies the gauge symmetry. Under the usual vertex transformation
\begin{equation}
    g_{ij}\longmapsto k_i g_{ij}k_j^{-1},\quad k_i\in G_{\sigma,\kappa},
\end{equation}
the insertion $\mathcal O_\Gamma^X[g]$ is not invariant under arbitrary $k_i\in G_{\sigma,\kappa}$. At the vertices of $\Gamma$, however, it is invariant under the stabilizer $H_\sigma$, since $H_\sigma$ leaves $x_0$ fixed. Thus the insertion reduces the vertex gauge symmetry from $G_{\sigma,\kappa}$ to $H_\sigma$ along $\Gamma$, while the full $G_{\sigma,\kappa}$ symmetry remains at the auxiliary vertices and is fixed by $\delta_{\rm pgf}[\gamma]$.

From this perspective, the Feynman graph may be viewed as a network of \emph{decorated Wilson lines} inserted in the BF path integral. Each edge $(ij)\in E_\Gamma$ carries the BF holonomy $g_{ij}$, but instead of the usual representation theoretic Wilson line insertion, such as a matrix element or a character, it is decorated by the propagator kernel
\begin{equation}
    W_{G_F}(g_{ij})
    :=
    G_F\!\left(
    d_X(x_0,g_{ij}x_0)
    \right).
\end{equation}
Hence
\begin{equation}
    \mathcal O_\Gamma^X[g]
    =
    \prod_{(ij)\in E_\Gamma}
    W_{G_F}(g_{ij}).
\end{equation}
The vertices of the Feynman graph are then precisely the points where these decorated open holonomies meet, with their residual $H_\sigma$ symmetry inherited from the homogeneous space description $X=G_{\sigma,\kappa}/H_\sigma$. In this sense, the original Feynman diagram is recovered inside the topological BF theory as a graph of Wilson line type defects whose holonomies are weighted by the scalar propagator as in \eqref{eq:scalar_propagator_as_wilson_sum}.

\section{Regge dilaton gravity from Feynman diagrams}\label{sec:regge-reconstruction}

In this section we henceforth restrict to the Euclidean ($\sigma = 1$) case and prove the two-dimensional analog of \cite{baratin_hidden_2007,baratin_hidden_2007-1} simultaneously in the flat $(\kappa = 0)$ and homogeneously curved $(\kappa = 1, \kappa = -1)$ cases. That is, we show that the reduced Feynman amplitude \eqref{eq:red-intro-feynman} can be re-cast into a Regge type path integral with operator insertion, giving the interpretation that the background geometry of the Feynman amplitude is generated by the dynamics of the overarching Regge theory. The Regge theory in question is the homogeneous Regge discretization of dilaton gravity of \cite{girelli_regge_2026}, with two modifications, both needed to reproduce the Feynman amplitude exactly: the deficit angle constraint is imposed only modulo $2\pi$, equivalently the dilaton is integer valued, and the triangles carry orientations which are summed over (see Remark \ref{rem:int-dilaton-ori-sum} for more details on the comparison with \cite{girelli_regge_2026}).

First we recall the definition and the geometric properties of homogeneously curved triangles. Then we show how the Feynman diagram amplitude measure can be rewritten in terms of relative (geodesic) lengths. We finally use that in general not all relative lengths are independent, which can be translated in terms of deficit angle properties, precisely encoding the topological model we are after. 

\subsection{Geometry of homogeneous triangles}\label{subsubsec:homogeneous-triangles}

The BF construction used group variables to encode relative positions, whereas to recover homogeneous Regge dilaton gravity we instead change from vertex position variables to $G_{1,\kappa}$-invariant lengths,
\begin{equation}
    l_{ij} := d_{X}(x_{i}, x_{j}), \: x_{i}, x_{j}\in X.\label{eq:regge-edge-lengths}
\end{equation}
Our goal is to build homogeneously curved triangles from the Feynman amplitude vertex position data. This subsection is thus dedicated to reviewing the basic notions of homogeneously curved Regge simplices.

To treat the $\kappa = 1, 0, -1$ cases simultaneously, it is convenient to introduce the curvature dependent functions,
\begin{equation}
    S_{\Lambda}(z):=\begin{cases}
        \frac{\sin(\sqrt{\Lambda}z)}{\sqrt{\Lambda}} ,\quad \Lambda >0 \\
        z, \quad \Lambda = 0\\
       \frac{\sinh(\sqrt{-\Lambda}z)}{\sqrt{-\Lambda}},\quad \Lambda <0
    \end{cases},\quad C_{\Lambda}(z):= \dv{z}S_{\Lambda}(z),\quad z>0,
\end{equation}
and, for every $\tau = (ijk)$, the Gram matrix
\begin{equation}
 G_{\tau,\Lambda}:= \begin{pmatrix}
        1 & C_{\Lambda}(l_{ij}) & C_{\Lambda}(l_{ik})\\
        C_{\Lambda}(l_{ij}) & 1 & C_{\Lambda}(l_{jk})\\
        C_{\Lambda}(l_{ik}) & C_{\Lambda}(l_{jk}) & 1
    \end{pmatrix}\label{eq:Gram-matrix}
    \end{equation}
and triangle Jacobian,
\begin{equation}
        V_{\tau,\Lambda}:=S_{\Lambda}(l_{ij})S_{\Lambda}(l_{ik})\sin\beta^{\tau}_{i}=\begin{cases}
            \frac{1}{\abs{\Lambda}}\sqrt{\mathrm{det}(G_{\tau,\Lambda})},\:\Lambda \neq 0 \\
            2\mathcal{V}_{\tau,0},\:\Lambda = 0
        \end{cases}\label{eq:triangle-jacobian}
\end{equation}
where $\beta^{\tau}_{i}\in (0,\pi)$ is the interior angle between $(ij)$ and $(ik)$, and $\mathcal{V}_{\tau, 0}$ is the Euclidean area of $\tau$. 

\paragraph{Geometry of the triangulation.} We say the edge length assignments to a triangle $\tau$ are \textit{admissible} if they satisfy the appropriate strict triangle inequalities. Spherical triangles additionally are required to have $l_{ij}<\pi/\sqrt{\abs{\Lambda}}$ and perimeter less than $2\pi/\sqrt{\abs{\Lambda}}$. All such $\tau$ with admissible length assignments thus satisfy $V_{\tau,\Lambda}>0$.

The generalized cosine law gives $\beta^{\tau}_{i}$ in terms of the relative edge lengths as,
\begin{equation}
 C_\Lambda(l_{jk})
 =C_\Lambda(l_{ij})C_\Lambda(l_{ik})
  +\Lambda S_\Lambda(l_{ij})S_\Lambda(l_{ik})\cos\beta_i^\tau,
\label{eq:generalized-cosine-law}
\end{equation}
where in the flat limit $\Lambda \to 0$ one recovers,
\begin{equation}
    l_{jk}^2=l_{ij}^2+l_{ik}^2-2l_{ij}l_{ik}\cos\beta_i^\tau.\label{eq:flat-cosine-law}
\end{equation}
 For $\Lambda\neq0$ we realize $X$ as the two-dimensional quadric
 \begin{equation}
    X_{\Lambda}:= \{\mathcal{X}\in\R^{3}\:|\:\Lambda\,\mathcal{X}\cdot \mathcal{X} = 1\}\label{eq:quadric}
 \end{equation}
 taking the upper sheet for $\Lambda<0$, where $\mathcal{X}\cdot \mathcal{X}'$ is the Euclidean inner product for $\Lambda>0$ and the Minkowski inner product for $\Lambda<0$. Two points $x_{i},x_{j}\in X$ with ambient vectors $\mathcal{X}_{i},\mathcal{X}_{j}$ thus satisfy
 \begin{equation}
     \Lambda\,\mathcal{X}_{i}\cdot \mathcal{X}_{j} = C_{\Lambda}(l_{ij}).\label{eq:quadric-inner-product}
 \end{equation}
 The vertices of a given $\tau = (123)$ are represented by three points,
\begin{equation}
    \mathcal{X}_1,\mathcal{X}_2,\mathcal{X}_3\in X_\Lambda,
    \quad
    \mathcal{X}_i\cdot \mathcal{X}_i=\frac{1}{\Lambda}.
\end{equation}
For $\Lambda = 0$ we set $\mathcal{X}_{i}:=(1,x_{i})\in\R^{3}$. A more detailed discussion of this quadric realization is available in \cite{girelli_regge_2026}.

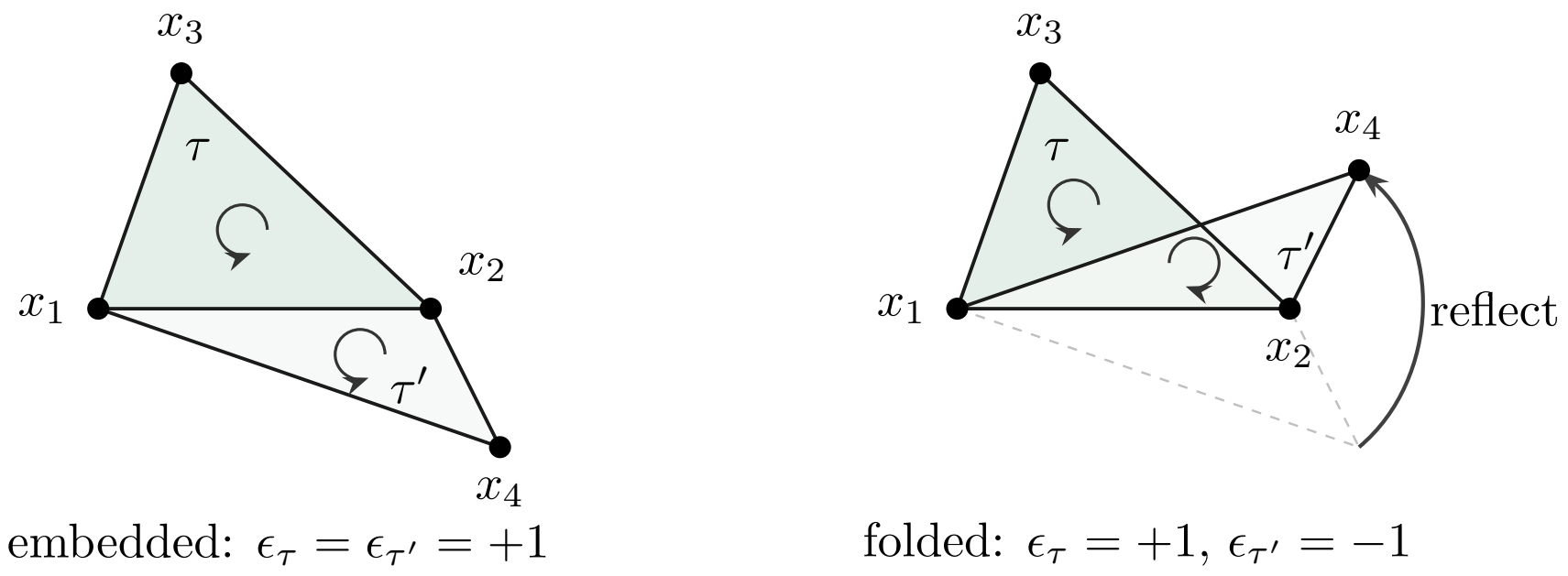
\begin{figure}[h]
\centering
\begin{tikzpicture}[scale=1.0,font=\small]
 \coordinate (A) at (0,0);   \coordinate (B) at (2.4,0);
 \coordinate (C) at (0.6,1.7); \coordinate (D) at (2.9,-1.0);
 \fill[jtSage] (A)--(B)--(C)--cycle;
 \fill[jtSage!35!white] (B)--(A)--(D)--cycle;
 \draw[jt edge] (A)--(B)--(C)--cycle;
 \draw[jt edge] (A)--(D)--(B);
 \draw[black!80,line width=.6pt,-{Stealth[length=1.8mm,width=1.4mm]}] ($(1.0,0.57)+(0.22,0)$) arc[start angle=0,end angle=290,radius=0.18];
 \draw[black!80,line width=.6pt,-{Stealth[length=1.8mm,width=1.4mm]}] ($(1.85,-0.33)+(0.22,0)$) arc[start angle=0,end angle=290,radius=0.18];
 \node[jt label] at (0.72,1.15) {$\tau$};
 \node[jt label] at (2.25,-0.55) {$\tau'$};
 \foreach \p/\lab/\pos in {A/1/left,B/2/above right,C/3/above,D/4/below}
   \node[vertex,label=\pos:$x_{\lab}$] at (\p) {};
 \node[align=center] at (1.3,-1.7) {embedded: $\epsilon_{\tau}=\epsilon_{\tau'}=+1$};
 \begin{scope}[xshift=6.2cm]
  \coordinate (A1) at (0,0);   \coordinate (B1) at (2.4,0);
  \coordinate (C1) at (0.6,1.7); \coordinate (D0) at (2.9,-1.0);
  \coordinate (D1) at (2.9,1.0);
  \draw[jt faint] (A1)--(D0)--(B1);
  \draw[jt move] (D0) to[bend right=50] node[right,jt label] {reflect} (D1);
  \fill[jtSage] (A1)--(B1)--(C1)--cycle;
  \fill[jtSage!35!white,opacity=.75] (B1)--(A1)--(D1)--cycle;
  \draw[jt edge] (A1)--(B1)--(C1)--cycle;
  \draw[jt edge] (A1)--(D1)--(B1);
  \draw[black!80,line width=.6pt,-{Stealth[length=1.8mm,width=1.4mm]}] ($(0.8,0.75)+(0.22,0)$) arc[start angle=0,end angle=290,radius=0.18];
  \draw[black!80,line width=.6pt,-{Stealth[length=1.8mm,width=1.4mm]}] ($(1.85,0.33)+(-0.32,0)$) arc[start angle=180,end angle=-110,radius=0.18];
  \node[jt label] at (0.72,1.15) {$\tau$};
  \node[jt label] at (2.45,0.42) {$\tau'$};
  \foreach \p/\lab/\pos in {A1/1/left,B1/2/below,C1/3/above,D1/4/above}
    \node[vertex,label=\pos:$x_{\lab}$] at (\p) {};
  \node[align=center] at (1.3,-1.7) {folded: $\epsilon_{\tau}=+1$, $\epsilon_{\tau'}=-1$};
 \end{scope}
\end{tikzpicture}
\caption{Meaning of the orientation sign \eqref{eq:orientation-sign}, drawn for $\Lambda =0$ and triangles $\tau = (123), \tau' = (214)$. Left: both cyclic orders agree with the orientation of $X$, $\epsilon_{\tau}=\epsilon_{\tau'}=+1$, and the two triangles lie on opposite sides of their common edge. Right: $\tau'$ is traversed in the opposite sense, $\epsilon_{\tau'}=-1$, i.e., it is folded over the edge $(12)$ onto the same side as $\tau$. Since edge lengths do not distinguish the two developments, both enter the measure, cf. Lemma~\ref{lem:one-point-insertion}, \eqref{eq:inv-measure-final} and \eqref{eq:oriented-Regge}.}
\label{fig:folded}
\end{figure}

\paragraph{Orientations and development.} We now fix our conventions for orientations, which play an important role below. The orientation of a triangle $\tau = (ijk)$ with vertices $x_{i},x_{j},x_{k}\in X$ is defined as
\begin{equation}
    \epsilon_{\tau}:=\mathrm{sgn}(\mathrm{det}(\mathcal{X}_{i},\mathcal{X}_{j},\mathcal{X}_{k}))\in\{\pm1\},\label{eq:orientation-sign}
\end{equation}
which is invariant under cyclic permutations of $(ijk)$, flips under transpositions, and equals $+1$ precisely when $(x_{i},x_{j},x_{k})$ is positively oriented in $X$. The determinant itself is the signed version of the triangle Jacobian \eqref{eq:triangle-jacobian},
\begin{equation}
    \mathrm{det}(\mathcal{X}_{i},\mathcal{X}_{j},\mathcal{X}_{k}) = \begin{cases}
        \epsilon_{\tau}\abs{\Lambda}^{-1/2}V_{\tau,\Lambda}, & \Lambda\neq0\\
        \epsilon_{\tau}V_{\tau,0}, & \Lambda = 0
    \end{cases}\label{eq:signed-jacobian}
\end{equation}
as follows from $\mathrm{det}(\mathcal{X}_{i},\mathcal{X}_{j},\mathcal{X}_{k})^{2} = \abs{\Lambda}^{-3}\mathrm{det}(G_{\tau,\Lambda})$ for $\Lambda\neq0$, and from $\mathrm{det}(\mathcal{X}_{i},\mathcal{X}_{j},\mathcal{X}_{k})$ being twice the signed Euclidean area for $\Lambda=0$. Every $\tau\in T(\Delta)$ inherits a cyclic order of its vertices from the orientation of the disk, and $\epsilon_{\tau}$ is always understood with respect to this order, thus $\epsilon_{\tau} = -1$ signals that the developed triangle is reflected (``folded'') relative to the abstract orientation (see Figure \ref{fig:folded}). 

 Two consequences of these definitions will be used repeatedly. Firstly, let $0,1,2,3\in X$ be four points and let $\tau_{i}:=(0123)\setminus\{i\}$ be the four triangles $(123)$, $(023)$, $(031)$ and $(012)$. Expanding $\mathcal{X}_{0}$ in the basis $\mathcal{X}_{1},\mathcal{X}_{2},\mathcal{X}_{3}$ by Cramer's rule and using \eqref{eq:signed-jacobian} we find
 \begin{align}
     \mathrm{det}(\mathcal{X}_{1}, \mathcal{X}_{2}, \mathcal{X}_{3})\mathcal{X}_{0} &= \mathrm{det}(\mathcal{X}_{0}, \mathcal{X}_{2}, \mathcal{X}_{3})\mathcal{X}_{1}+ \mathrm{det}(\mathcal{X}_{1}, \mathcal{X}_{0}, \mathcal{X}_{3})\mathcal{X}_{2}+ \mathrm{det}(\mathcal{X}_{1}, \mathcal{X}_{2}, \mathcal{X}_{0})\mathcal{X}_{3}\nonumber\\
     \implies &\epsilon_{\tau_{0}}V_{\tau_{0},\Lambda}\mathcal{X}_{0}=\epsilon_{\tau_{1}}V_{\tau_{1},\Lambda}\mathcal{X}_{1}+\epsilon_{\tau_{2}}V_{\tau_{2},\Lambda}\mathcal{X}_{2}+\epsilon_{\tau_{3}}V_{\tau_{3},\Lambda}\mathcal{X}_{3}.\label{eq:cramers}
 \end{align}
 Contracting both sides with $\Lambda \mathcal{X}_{0}$, using \eqref{eq:quadric} and \eqref{eq:quadric-inner-product} gives the signed Jacobian identity\footnote{Recalling that we take $\mathcal{X}_{i}:=(1,x_{i})\in\R^{3}$ for $\Lambda = 0$, and \eqref{eq:triangle-jacobian} for the Jacobian factor, the desired identity in the flat case is just the first component of \eqref{eq:cramers}, giving,
 \begin{equation*}
     \epsilon_{0}\mathcal{V}_{0} = \sum_{i=1}^{3}\epsilon_{i}\mathcal{V}_{i},\quad \mathcal{V}_{i} := \mathcal{V}_{\tau_{i},0}, \quad \epsilon_{i} := \epsilon_{\tau_{i}}
 \end{equation*}
 which, since $C_{0}(l_{0i}) = 1$, agrees with the $\Lambda = 0$ limit of \eqref{eq:signed-jacobian-identity}.}
\begin{equation}
    \epsilon_{0}V_{0} = \sum_{i=1}^{3}\epsilon_{i}C_{\Lambda}(l_{0i})V_{i},\quad V_{i}:=V_{\tau_{i},\Lambda},\quad \epsilon_{i}:=\epsilon_{\tau_{i}}.\label{eq:signed-jacobian-identity}
\end{equation}
For the vertex $0$ inside the triangle $(123)$ and all orientations positive, the above states that the Jacobian of $(123)$ is the sum of those of its three sub-triangles, up to the curvature corrections $C_{\Lambda}(l_{0i})$. For $\Lambda = 0$ it reduces to the additivity of signed areas. 

Secondly, if $v$ is an interior vertex of $\Delta$ and the triangles $\tau\ni v$ are developed in $X$ around $x_{v}$, the signed angles $\epsilon_{\tau}\beta^{\tau}_{v}$ add up to an integer multiple of $2\pi$,
\begin{equation}
    \sum_{\tau\ni v}\epsilon_{\tau}\beta^{\tau}_{v} = 2\pi w_{v},\quad w_{v}\in\Z,\label{eq:angular-closure}
\end{equation}
because the direction of the geodesic from $x_{v}$ to each successive neighboring vertex is obtained from the previous one by a rotation through the signed angle of the developed triangle in between, and because after a full turn around $v$ one returns to the initial direction. The integer $w_{v}$ is the winding number of the developed vertex face of $v$ around $x_{v}$, equaling one for an embedded, consistently oriented dual face, while $w_{v}\neq1$ characterizes the branched configurations discussed in \cite{girelli_regge_2026}, as well as configurations with folded triangles, which were excluded in \cite{girelli_regge_2026}.

\subsection{Lifting vertex positions to relative lengths}

We begin with re-writing the invariant measure $d\mu_{X}(\bullet)$ in terms of the relative lengths \eqref{eq:regge-edge-lengths}, adapting the methods of \cite{baratin_hidden_2007,baratin_hidden_2007-1} to our two-dimensional setting. 

A useful preliminary question to ask is how many degrees of freedom need to be integrated over to achieve this goal. We note that since $\Gamma$ is made up of $N$ vertices and the isometry group leaving \eqref{eq:intro-feynman} invariant is three-dimensional, we expect $2N-3$ integrations. In terms of the relative lengths between $N$ vertices, we have $N(N-1)/2$ variables, however not all are independent. In general there are 
\begin{equation}
    \frac{1}{2}N(N-1)-(2N-3)=\frac{1}{2}(N-2)(N-3)\label{eq:num-redundant-lengths},
\end{equation}
\textit{redundant} relative lengths. We denote such redundant relative lengths by $l'^{\epsilon_{\tau}}_{ab}(l)$. In general, as the notation suggests, the geometry of $\Gamma$ allows such redundant relative lengths to be reconstructed from the relevant independent relative lengths and the relevant orientations. We will see later that the \textit{redundancy} of the $l'^{\epsilon_{\tau}}_{ab}(l)$ is closely related to the geometry of $\Gamma$, hence, by ``de-parameterizing'' the background geometry we may lift the $l'^{\epsilon_{\tau}}_{ab}(l)$ to be dynamical, introducing a Regge deficit angle constraint.

We illustrate the basic idea of this re-writing of $d\mu_{X}(\bullet)$ first in the case of one triangle, constructed by considering the possible ways to add a single vertex to a pair of vertices. Our final re-writing will be achieved by working recursively.

\begin{figure}[ht]
\centering
\begin{tikzpicture}[scale=0.95,
    orient/.style={draw=black!80,line width=.7pt,-{Stealth[length=1.8mm,width=1.4mm]}},
    angle/.style={draw=black!70,line width=.55pt,-{Stealth[length=1.3mm,width=1mm]}}]
 \begin{scope}
  \coordinate (x1) at (0,0); \coordinate (x2) at (3.2,0);
  \draw[jt faint] (-0.4,0)--(3.6,0);
  \draw[jt edge] (x1)--(x2);
  \node[jt label,font=\footnotesize,anchor=south] at (1.6,0.05) {$l_{12}$};
  \node[vertex,label={[jt label]below:$x_1$}] at (x1) {};
  \node[vertex,label={[jt label]below:$x_2$}] at (x2) {};
 \end{scope}
 \draw[jt move] (4.0,0.12)--(5.9,1.0);
 \draw[jt move] (4.0,-0.12)--(5.9,-1.0);
 \node[font=\footnotesize] at (5.0,0) {insert $y$};
 \begin{scope}[xshift=6.6cm]
  \begin{scope}[yshift=1.0cm]
   \coordinate (x1) at (0,0); \coordinate (x2) at (3.2,0); \coordinate (y) at (1.1,1.7);
   \draw[jt faint] (-0.4,0)--(3.6,0);
   \fill[jtSage] (x1)--(x2)--(y)--cycle;
   \draw[jt edge] (x1)--(x2)--(y)--cycle;
   \draw[angle] (0.5,0) arc (0:57:0.5);
   \node[jt label,font=\footnotesize] at (20:0.98) {$\beta^{\tau}_{1y}$};
   \draw[orient] (1.85,0.6) ++(-60:0.25) arc (-60:240:0.25);
   \node[jt label,font=\footnotesize,anchor=north] at (1.6,-0.05) {$l_{12}$};
   \node[jt label,font=\footnotesize,anchor=east] at (0.5,0.92) {$l_{1y}$};
   \node[jt label,font=\footnotesize,anchor=west] at (2.2,0.92) {$l_{2y}$};
   \node[vertex,label={[jt label]below:$x_1$}] at (x1) {};
   \node[vertex,label={[jt label]below:$x_2$}] at (x2) {};
   \node[vertex,label={[jt label]above:$y$}] at (y) {};
   \node[font=\small,anchor=west] at (3.4,0.9) {$\epsilon_\tau=+1$};
  \end{scope}
  \begin{scope}[yshift=-1.0cm]
   \coordinate (x1) at (0,0); \coordinate (x2) at (3.2,0); \coordinate (y) at (1.1,-1.7);
   \draw[jt faint] (-0.4,0)--(3.6,0);
   \fill[jtSage!15!white] (x1)--(x2)--(y)--cycle;
   \draw[jt edge] (x1)--(x2)--(y)--cycle;
   \draw[angle] (0.5,0) arc (0:-57:0.5);
   \node[jt label,font=\footnotesize] at (-18:1.06) {$-\beta^{\tau}_{1y}$};
   \draw[orient] (1.85,-0.6) ++(60:0.25) arc (60:-240:0.25);
   \node[jt label,font=\footnotesize,anchor=south] at (1.6,0.05) {$l_{12}$};
   \node[jt label,font=\footnotesize,anchor=east] at (0.5,-0.92) {$l_{1y}$};
   \node[jt label,font=\footnotesize,anchor=west] at (2.2,-0.92) {$l_{2y}$};
   \node[vertex,label={[jt label]above:$x_1$}] at (x1) {};
   \node[vertex,label={[jt label]above:$x_2$}] at (x2) {};
   \node[vertex,label={[jt label]below:$y$}] at (y) {};
   \node[font=\small,anchor=west] at (3.4,-0.9) {$\epsilon_\tau=-1$};
  \end{scope}
  \end{scope}
\end{tikzpicture}
\caption{One-point insertion (Lemma~\ref{lem:one-point-insertion}), drawn for $\Lambda=0$. Given $x_1,x_2$ and the lengths $l_{1y},l_{2y}$, the possible insertions $y$ lie on one of the two sides of the geodesic through $x_1$ and $x_2$. The two possible triangles (dark green and light green) $\tau=(x_1x_2y)$ have the same edge lengths but opposite orientations depending on which side $y$ is inserted, and corresponding to the two signs of the polar angle $\pm\beta^\tau_{1y}$ of $y$ about $x_1$.}
\label{fig:one-point-insertion}
\end{figure}
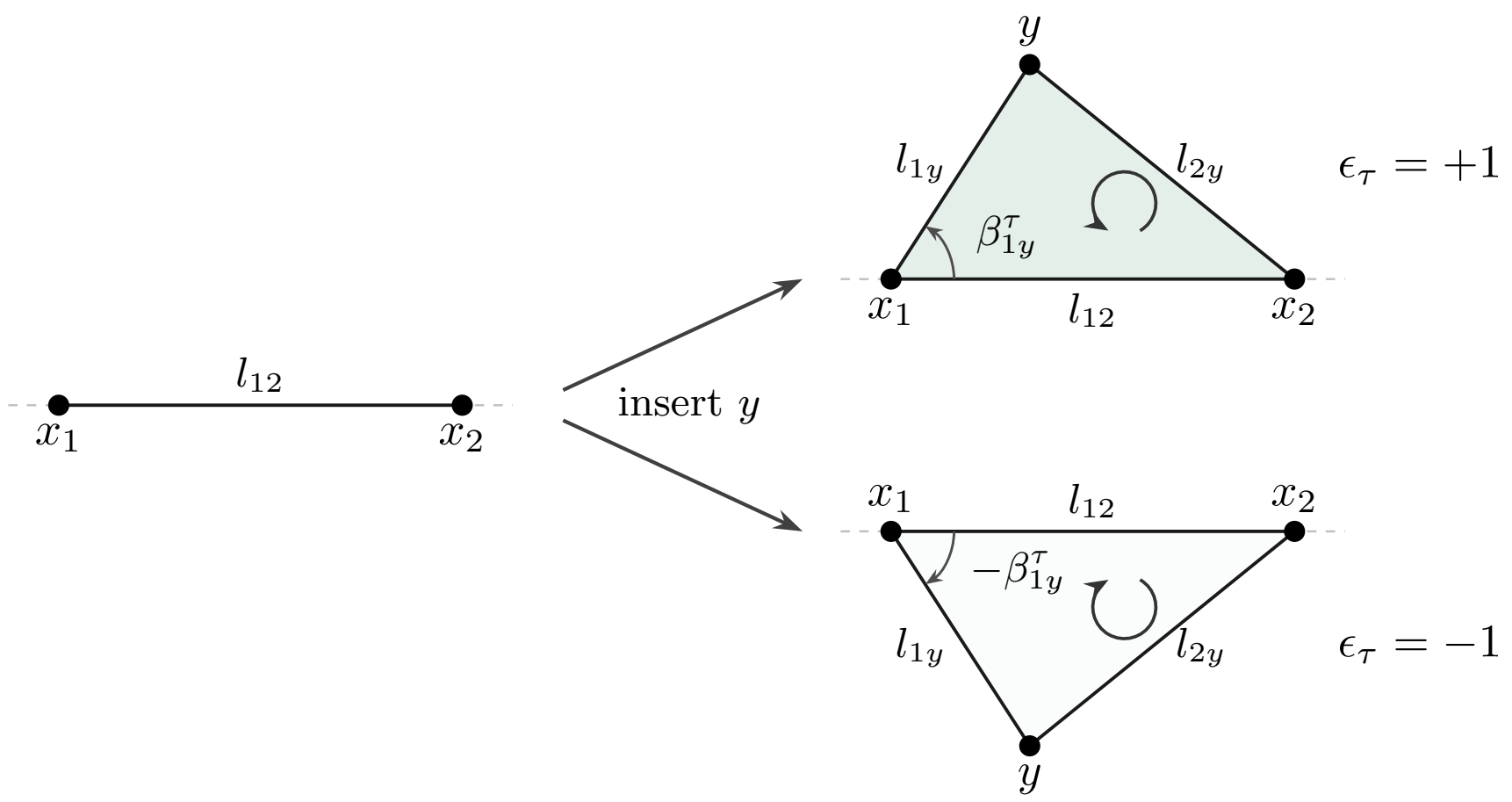

\begin{lemma}[One-point insertion]\label{lem:one-point-insertion}
    Fix two distinct points $x_{1}, x_{2}\in X$ (non-antipodal in the spherical case) such that $d_{X}(x_{1},x_{2}):=l_{12} >0$, and let $l_{1y} := d_{X}(x_{1},y)$, $l_{2y} := d_{X}(x_{2},y)$ for some $y\in X$. Let $\tau = (x_{1}x_{2}y)$ be the corresponding oriented geodesic triangle, then on the non-degenerate sector we have,
    \begin{equation}
        d\mu_{X}(y) = \sum_{\epsilon_{\tau} = \pm 1}\frac{S_{\Lambda}(l_{1y})dl_{1y} S_{\Lambda}(l_{2y})dl_{2y}}{V_{\tau,\Lambda}}
    \end{equation}
    where $\epsilon_{\tau} = \pm 1$ corresponds to the two possible orientations of $\tau$ (see Figure \ref{fig:one-point-insertion}).
\end{lemma}
\begin{proof}
   Here $N= 3$, thus from \eqref{eq:num-redundant-lengths} there are no redundant relative lengths. We use geodesic polar coordinates $(l_{1y},\beta_{1y}^{\tau})$ around $x_{1}$, where $\beta_{1y}^{\tau}$ is the internal angle measured from the geodesic joining $x_{1}$ to $x_{2}$, given by the homogeneous space cosine law,
    \begin{equation}
        \cos(\beta_{1y}^{\tau}) = \frac{C_{\Lambda}(l_{2y})- C_{\Lambda}(l_{1y})C_{\Lambda}(l_{12})}{\Lambda S_{\Lambda}(l_{1y})S_{\Lambda}(l_{12})}.
    \end{equation}
    In these coordinates the invariant measure $d\mu_{X}(y)$ admits the decomposition,
    \begin{equation}
        d\mu_{X}(y) = S_{\Lambda}(l_{1y})dl_{1y}\:d\beta^{\tau}_{1y}.\label{eq:inv-measure-part1}
    \end{equation}
    By differentiating \eqref{eq:generalized-cosine-law} (for $\Lambda = 0$, the flat cosine law \eqref{eq:flat-cosine-law}, here and above) with respect to $l_{2y}$, holding $l_{1y}$ and $l_{12}$ fixed, we can re-express $d\beta^{\tau}_{1y}$ as,
    \begin{equation}
        S_{\Lambda}(l_{2y})|dl_{2y}| = S_{\Lambda}(l_{1y})S_{\Lambda}(l_{12})|\sin(\beta_{1y}^{\tau})||d\beta_{1y}^{\tau}| = V_{\tau,\Lambda}|d\beta_{1y}^{\tau}| \implies |d\beta^{\tau}_{1y}|  = \frac{S_{\Lambda}(l_{2y})|dl_{2y}|}{V_{\tau,\Lambda}}.
    \end{equation}
     The sign of $\beta^{\tau}_{1y}$ depends on which side of the initial edge $(12)$ that the inserted vertex $y$ lies on. That is, the insertion of $y$ can create two possible triangles $(12y)$, each with opposite orientations. Plugging this back into \eqref{eq:inv-measure-part1} for these possible signs gives the sum over the two possible orientations, yielding the claimed expression.
\end{proof}

An immediate consequence of this result is that for a single triangle we have,
\begin{equation}
 \frac{\prod_{i=1}^3d\mu_X(x_i)}{\Vol(G_{1,\kappa})}
 =\sum_{\epsilon_\tau=\pm1}
   \frac{\prod_{e\subset\tau}S_\Lambda(l_e)d l_e}
        {V_{\tau,\Lambda}}.
 \label{eq:triangle-measure}
\end{equation}
To extend these results beyond a single triangle to our generic $\Delta$ of the disk, we first treat the simpler case of a triangulation with no internal vertices.

Let $\Delta_{k}$ be a triangulation with homogeneously curved triangles of the disk $D$ with $N = 2+k$ vertices, constructed by beginning with one edge, such that at each step we choose a boundary edge and glue one new triangle along it, thereby introducing one new boundary vertex. Such a complex satisfies
\begin{equation}
    \abs{T(\Delta_{k})} = N-2,\quad |E(\Delta_{k})| = 2N-3,\quad \abs{V^{0}(\Delta_{k})} = 0,
\end{equation}
where $V^{0}(\Delta_{k})$ denotes the collection of internal vertices, meaning $\Delta_{k}$ is essentially a polygon. No embedding requirement is imposed in this development procedure. In particular an abstract new boundary vertex may develop on either side of its attachment edge. The sign $\epsilon_\tau=\pm1$ records the orientation of the developed triangle relative to the chosen orientation of the abstract disk.

Note that the number of edges equals the number of continuous degrees of freedom of $N$ points in a two-dimensional homogeneous space after quotienting by its three-dimensional isometry group. By construction we have $V(\Delta_{k}) = V_{\Gamma}$, and three ``types'' of edges: edges in $E_{\Gamma}\cap E(\Delta_{k})$, edges in $E(\Delta_{k})\setminus E_{\Gamma}$, and edges in $E_{\Gamma}\setminus E(\Delta_{k})$. The first type of edges are those whose associated relative length will appear as independent integration variables as well as in the scalar matter propagator, the second type of edges are those whose relative length will appear as independent integration variables but have no matter propagator, and the third type of edges are those whose associated relative length is redundant.

\begin{proposition}[Recursive invariant-distance measure]\label{prop:polygon-measure}
    Let $x_{1}, \ldots, x_{N}\in X$ lie in the non-degenerate sector associated with $\Delta_{k}$. Then,
    \begin{equation}
       \frac{1}{\mathrm{Vol}(G_{1,\kappa})} \prod_{i=1}^{N}d\mu_{X}(x_{i}) = \sum_{\{\epsilon_{\tau} = \pm 1\}}\prod_{e\in E(\Delta_{k})}S_{\Lambda}(l_{e})dl_{e}\prod_{\tau\in T(\Delta_{k})}\frac{1}{V_{\tau,\Lambda}}.\label{eq:inv-measure-final}
    \end{equation}
\end{proposition}
\begin{proof}
   We proceed with a proof by induction. For two points $x_{1},x_{2}\in X$ with $l_{12}:= d_{X}(x_{1}, x_{2})$, the map from $(g,l_{12})$ to the ordered pair $(x_{1}, x_{2})$ has Jacobian $S_{\Lambda}(l_{12})$ with $dg$ normalized as in \eqref{eq:haar-normalization},
    \begin{equation}
        d\mu_{X}(x_{1})d\mu_{X}(x_{2}) = dg S_{\Lambda}(l_{12}) dl_{12}.
    \end{equation}
    Indeed, writing $d\mu_{X}(x_{2}) = S_{\Lambda}(l_{12})dl_{12}d\beta$ in geodesic polar coordinates around $x_{1}$, the position $x_{1}$ and the direction $\beta$ of the geodesic from $x_{1}$ to $x_{2}$ are precisely the data of a group element $g\in G_{1,\kappa}$, with $dg = d\mu_{X}(x_{1})d\beta$ as in \eqref{eq:haar-normalization}.
    
   By Lemma \ref{lem:one-point-insertion}, inserting a third point $x_{3}\in X$ gives,
    \begin{align}
        d\mu_{X}(x_{1})d\mu_{X}(x_{2})d\mu_{X}(x_{3}) &= dg S_{\Lambda}(l_{12})dl_{12}\sum_{\epsilon_{(123)} = \pm 1}\frac{S_{\Lambda}(l_{13})dl_{13}S_{\Lambda}(l_{23})dl_{23}}{V_{(123),\Lambda}}\nonumber\\
        \implies \frac{1}{\mathrm{Vol}(G_{1,\kappa})}\prod_{i=1}^{3}d\mu_{X}(x_{i}) &= \sum_{\epsilon_{(123)} = \pm 1}\frac{S_{\Lambda}(l_{12})dl_{12}S_{\Lambda}(l_{13})dl_{13}S_{\Lambda}(l_{23})dl_{23}}{V_{(123),\Lambda}}
    \end{align}
    which is our base case for the induction (i.e., \eqref{eq:inv-measure-final} for $k=1$). 
    
    Assume now that \eqref{eq:inv-measure-final} holds for $\Delta_{k-1}$, which by definition contains $N= k+1$ vertices. Let $\Delta_{k}$ be obtained from $\Delta_{k-1}$ by gluing a new triangle $\tau = (abw)$ along the boundary edge $(ab)$, with new vertex $w$, so that $N = k+2$. Since $x_{a}$ and $x_{b}$ are already present, Lemma \ref{lem:one-point-insertion} applied to $y = x_{w}$ gives
    \begin{equation}
        d\mu_{X}(x_{w}) = \sum_{\epsilon_{\tau} = \pm 1}\frac{S_{\Lambda}(l_{aw})dl_{aw} S_{\Lambda}(l_{bw})dl_{bw}}{V_{\tau,\Lambda}},
    \end{equation}
    which is precisely the contribution of the two new edges, the new inverse triangle factor and the new orientation sum to \eqref{eq:inv-measure-final} for $\Delta_{k}$. Thus by induction we have proven the Proposition.
\end{proof}
With the measure of \eqref{eq:intro-feynman} now expressed in terms of the relative lengths, we may re-write the entire reduced amplitude \eqref{eq:red-intro-feynman} similarly in terms of the relative lengths by noting integration of $\eqref{eq:inv-measure-final}$ over $G_{1,\kappa}$ gives the isometry group volume factor in the definition of the reduced amplitude. We thus find,
\begin{equation}
    I_{\Gamma}^{X,\mathrm{red}} = \int\prod_{e\in E(\Delta_{k})}S_{\Lambda}(l_{e})dl_{e}\sum_{\{\epsilon_{\tau} = \pm 1\}}\prod_{\tau\in T(\Delta_{k})}\frac{1}{V_{\tau,\Lambda}}\prod_{e\in E_{\Gamma}\cap E(\Delta_{k})}G_{F}(l_{e})\prod_{(ab)\in E_{\Gamma}\setminus E(\Delta_{k})}G_{F}(l'^{\epsilon_{\tau}}_{ab}(l)),\label{eq:relative-length-Feynman-amp}
\end{equation}
where as before $l'^{\epsilon_{\tau}}_{ab}(l_{\bullet})$ denotes every relative distance not belonging to $E(\Delta_{k})$, reconstructed from the edge lengths and the possible orientations.

\subsection{The recovery theorem}

To massage \eqref{eq:relative-length-Feynman-amp} into a Regge form we state and prove the two-dimensional homogeneous space version of \cite[equation (14)]{baratin_hidden_2007}. The idea is that the determination of the redundant relative lengths $l'^{\epsilon_{\tau}}_{ab}(l)$ from the independent relative lengths can be re-cast into a condition on the embeddability of simplices, which in turn will be shown to be related to the $1\leftrightarrow 3$ Pachner move  and a homogeneous space deficit angle constraint. This will allow us to extend the results of Proposition \ref{prop:polygon-measure} from a polygonal triangulation $\Delta_{k}$ to our general disk triangulation $\Delta$.

Consider four vertices $0,1,2,3$ with relative distances $l_{ij}$, and the associated four by four Gram matrix $G_{(0123),\Lambda}$ defined by,
\begin{equation}
   (G_{(0123),\Lambda})_{ij} = C_{\Lambda}(l_{ij}), \quad (G_{(0123),\Lambda})_{ii} = 1.
\end{equation}
Holding every $l_{ij}$ fixed except $l_{03}$ in $G_{(0123),\Lambda}$, denote the corresponding (normalized) determinant by,
\begin{equation}
D_{\Lambda}(l_{03}) := \begin{cases}
    \frac{1}{\Lambda^{3}}\mathrm{det}(G_{(0123),\Lambda}) ,\:\Lambda \neq 0\\
\frac{1}{8}\mathrm{det}(\mathrm{CM}_{(0123)}),\:\Lambda =0
\end{cases}   
\end{equation}
where $\mathrm{CM}_{(0123)}$ is the five by five Cayley--Menger matrix of the four points, whose first row and column are $(0,1,1,1,1)$ and whose remaining four by four block has entries $l_{ij}^{2}$, $i,j\in\{0,1,2,3\}$, with zeros on the diagonal. The normalization is chosen such that $D_{0}$ is the limit of $D_{\Lambda}$ as $\Lambda\to0$. In both cases $D_{\Lambda}$ vanishes exactly when the ambient vectors $\mathcal{X}_{i}$ of \eqref{eq:orientation-sign} are linearly dependent, i.e., the four vertices belong to the two-dimensional homogeneous space $X$ exactly when the associated Gram (respectively Cayley--Menger) matrix is degenerate, thus the geometry formed by this four vertex simplex is embeddable in $X$ precisely when $D_{\Lambda}(l_{03}) = 0$. 

Let us define four triangles $\tau_{i}$ by,
\begin{equation}
    \tau_{i}:=(0123)\setminus \{i\}.
\end{equation}
When $D_{\Lambda}(l_{03}) = 0$, $l_{03}$ can take on two possible values, reconstructed from the other independent relative lengths. For instance these two values depend on if the vertex $3$ of $\tau_{0}$ and the vertex $0$ of $\tau_{3}$ (which by definition share the edge $(12)$) lie on the same or opposite sides of $(12)$. Let us label these two possible roots of $D_{\Lambda}(l_{03})$ by $l'^{\epsilon_{0}\epsilon_{3}}_{03}$, where $\epsilon_i$ is the orientation \eqref{eq:orientation-sign} of $\tau_i$ (with the cyclic orders of \S\ref{subsubsec:homogeneous-triangles}), and define the (oriented) Regge deficit angle
\begin{equation}
 \delta_0^\epsilon(l;\Lambda)
 =2\pi-\sum_{i=1}^{3}\epsilon_i\beta_0^{\tau_i}(l;\Lambda).
 \label{eq:oriented-deficit-four}
\end{equation}
For all positive orientations this is the same definition as in \cite{girelli_regge_2026}, meaning that the above is an extension to allow for folded configurations. The criterion for reconstructing $l'^{\epsilon_{0}\epsilon_{3}}_{03}$ is captured by the following lemmas.

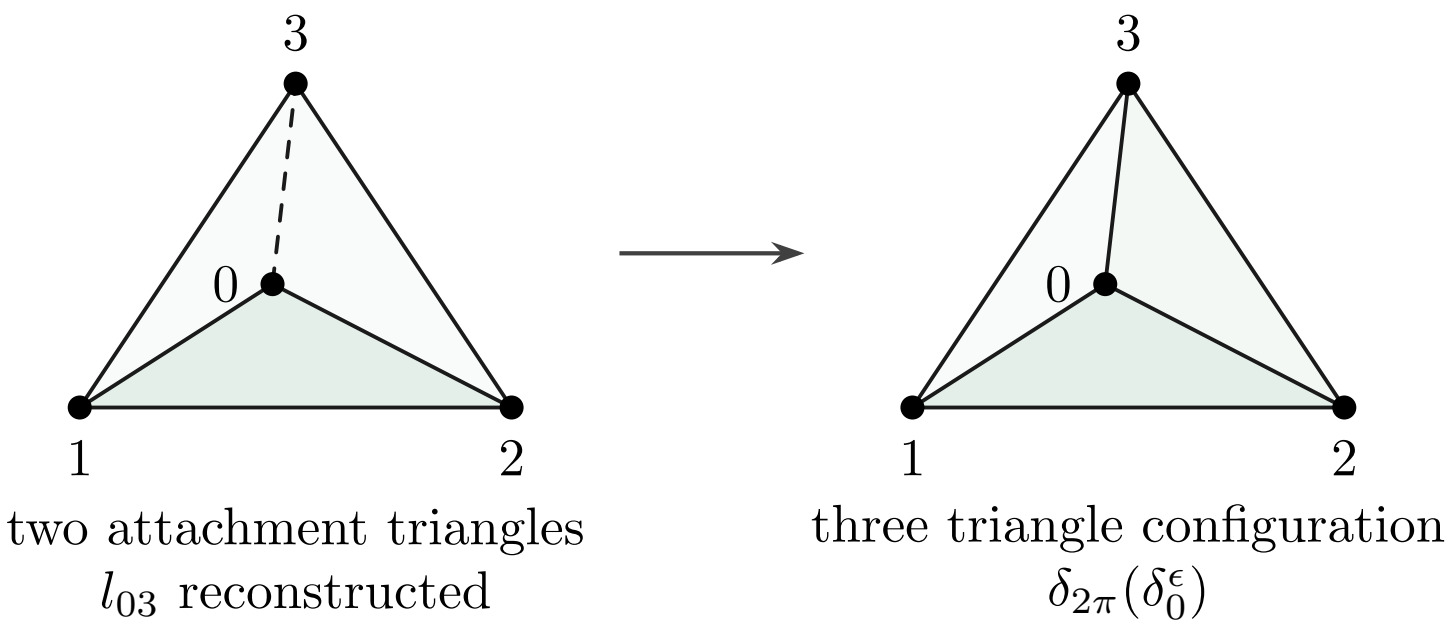
\begin{figure}[h]
\centering
\begin{tikzpicture}[scale=1.02,font=\small]
 \coordinate (A) at (0,0); \coordinate (B) at (2.8,0);
 \coordinate (C) at (1.4,2.1); \coordinate (O) at (1.25,.8);
 \fill[jtSage!25!white] (A)--(B)--(C)--cycle;
 \fill[jtSage] (A)--(B)--(O)--cycle;
 \draw[jt edge] (A)--(B)--(C)--cycle;
 \draw[jt edge] (A)--(O)--(B);
 \draw[jt edge,dashed] (O)--(C);
 \foreach \p/\lab/\pos in {A/1/below,B/2/below,C/3/above,O/0/left}
   \node[vertex,label=\pos:$\lab$] at (\p) {};
 \node[align=center] at (1.4,-1.0) {two attachment triangles\\
 $l_{03}$ reconstructed};
 \draw[jt move] (3.5,1)--(4.7,1);
 \begin{scope}[xshift=5.4cm]
  \coordinate (A1) at (0,0); \coordinate (B1) at (2.8,0);
  \coordinate (C1) at (1.4,2.1); \coordinate (O1) at (1.25,.8);
  \fill[jtSage] (A1)--(B1)--(O1)--cycle;
  \fill[jtSage!45!white] (B1)--(C1)--(O1)--cycle;
  \fill[jtSage!25!white] (C1)--(A1)--(O1)--cycle;
  \draw[jt edge] (A1)--(B1)--(C1)--cycle;
  \draw[jt edge] (A1)--(O1)--(B1) (O1)--(C1);
  \foreach \p/\lab/\pos in {A1/1/below,B1/2/below,C1/3/above,O1/0/left}
    \node[vertex,label=\pos:$\lab$] at (\p) {};
  \node[align=center] at (1.4,-1.0) {three triangle configuration\\
  $\delta_{2\pi}(\delta_0^\epsilon)$};
 \end{scope}
\end{tikzpicture}
\caption{The identity \eqref{eq:missing-length-replacement} for four points of $X$. Left: the triangles $(123)$ and $(012)$ share the edge $(12)$ and lie on the same side of it, so that $l_{03}$ (dashed) is determined by the other five lengths. Right: the same four points as a three triangle configuration with interior vertex $0$, where $l_{03}$ is integrated over and fixed by the periodic deficit angle constraint. Shown is the branch with $0$ inside $(123)$. The identity sums over both roots of $l_{03}$ and over all orientation branches, including folded ones.}
\label{fig:four-point}
\end{figure}

\begin{lemma}\label{lem:Gram-det-derivative}
    At a root of $D_{\Lambda}(l_{03}) = 0$, one has,
    \begin{equation}
        \abs{\pdv{D_{\Lambda}(l_{03})}{l_{03}}} = 2S_{\Lambda}(l_{03})V_{0}V_{3},\quad V_{i}:= V_{\tau_{i},\Lambda}.
    \end{equation}
\end{lemma}
\begin{proof}
    The proof is by direct computation using the definition of $D_{\Lambda}(l_{03})$. We have that $l_{03}$ shows up only in the $(0,3)$ and $(3,0)$ entries of $G_{(0123),\Lambda}$, hence giving,
    \begin{equation}
         \abs{\pdv{D_{\Lambda}(l_{03})}{l_{03}}} = \frac{2}{\abs{\Lambda}^{2}}S_{\Lambda}(l_{03})\abs{\mathrm{cof}_{03}},
    \end{equation}
    where $\mathrm{cof}_{03}$ denotes the corresponding cofactor of $G_{(0123),\Lambda}$. At a root of $D_{\Lambda}(l_{03}) = 0$, the cofactors obey the relations,
    \begin{equation}
        \abs{\mathrm{cof}_{03}}^{2} = \mathrm{cof}_{00}\mathrm{cof}_{33},
    \end{equation}
    and in general we have
    \begin{equation}
        \mathrm{cof}_{00} = \abs{\Lambda}^{2}V_{0}^{2},\quad  \mathrm{cof}_{33} = \abs{\Lambda}^{2}V_{3}^{2},
    \end{equation}
    therefore,
    \begin{equation}
        \abs{\mathrm{cof}_{03}} = \abs{\Lambda}^{2}V_{0}V_{3} .
    \end{equation}
    Plugging this back into our previous expression for the derivative we find the desired equality
    \begin{equation}
         \abs{\pdv{D_{\Lambda}(l_{03})}{l_{03}}} = 2S_{\Lambda}(l_{03})V_{0}V_{3}
    \end{equation}
    as claimed. For $\Lambda = 0$ the same computation applies to the Cayley--Menger form of $D_{0}$. The derivative of $l_{03}^{2}$ supplies the factor $2l_{03} = 2S_{0}(l_{03})$, and the diagonal cofactors are now $\mathrm{cof}_{00} = -4V_{0}^{2}$ and $\mathrm{cof}_{33} = -4V_{3}^{2}$.
\end{proof}
\begin{lemma}\label{lem:deficit-derivative}
    On the set $\delta_{0}^{\epsilon}(l;\Lambda) = 0$ (mod $2\pi$), for every compatible choice of orientations $\epsilon$, we have,
    \begin{equation}
        \abs{\pdv{\delta_{0}^{\epsilon}(l;\Lambda)}{l_{03}}}=  \frac{S_{\Lambda}(l_{03})V_{0}}{V_{1}V_{2}}.
    \end{equation}
\end{lemma}
\begin{proof}
    Again the proof follows from direct computation. 
    
    Only the triangles $\tau_{1} = (023)$ and $\tau_{2} = (031)$ contain the edge $(03)$, so only $\beta^{\tau_{1}}_{0}$ and $\beta^{\tau_{2}}_{0}$ depend on $l_{03}$. The homogeneous space cosine law \eqref{eq:generalized-cosine-law} gives,
    \begin{subequations}
    \begin{align}
        C_{\Lambda}(l_{23}) &= C_{\Lambda}(l_{02})C_{\Lambda}(l_{03})+\Lambda S_{\Lambda}(l_{02})S_{\Lambda}(l_{03})\cos(\beta_{0}^{\tau_{1}}),\\
         C_{\Lambda}(l_{13}) &= C_{\Lambda}(l_{01})C_{\Lambda}(l_{03})+\Lambda S_{\Lambda}(l_{01})S_{\Lambda}(l_{03})\cos(\beta^{\tau_{2}}_{0}).
    \end{align}
    \end{subequations}
    Differentiating both of the above with respect to $l_{03}$ we find,
    \begin{subequations}
    \begin{align}
        \pdv{\beta^{\tau_{1}}_{0}}{l_{03}} &= \frac{S_{\Lambda}(l_{02})C_{\Lambda}(l_{03})\cos(\beta^{\tau_{1}}_{0})-C_{\Lambda}(l_{02})S_{\Lambda}(l_{03})}{V_{1}},\\
        \pdv{\beta^{\tau_{2}}_{0}}{l_{03}} &= \frac{S_{\Lambda}(l_{01})C_{\Lambda}(l_{03})\cos(\beta^{\tau_{2}}_{0})-C_{\Lambda}(l_{01})S_{\Lambda}(l_{03})}{V_{2}}.
    \end{align}
    \end{subequations}
     Using $V_{1} = S_{\Lambda}(l_{02})S_{\Lambda}(l_{03})\sin\beta^{\tau_{1}}_{0}$, $V_{2} = S_{\Lambda}(l_{03})S_{\Lambda}(l_{01})\sin\beta^{\tau_{2}}_{0}$, $V_{3} = S_{\Lambda}(l_{01})S_{\Lambda}(l_{02})\sin\beta^{\tau_{3}}_{0}$, together with the constraint $\delta_{0}^{\epsilon} = 0$ (mod $2\pi$), one finds
    \begin{equation}
        V_{1}V_{2}\Big(\epsilon_{1}\pdv{\beta^{\tau_{1}}_{0}}{l_{03}}+ \epsilon_{2}\pdv{\beta^{\tau_{2}}_{0}}{l_{03}}\Big) = -\epsilon_{1}\epsilon_{2}S_{\Lambda}(l_{03})\Big(\epsilon_{1}C_{\Lambda}(l_{01})V_{1}+\epsilon_{2}C_{\Lambda}(l_{02})V_{2}+\epsilon_{3}C_{\Lambda}(l_{03})V_{3}\Big) = -\epsilon_{0}\epsilon_{1}\epsilon_{2}S_{\Lambda}(l_{03})V_{0}
    \end{equation}
   where the last step is the signed Jacobian identity \eqref{eq:signed-jacobian-identity}. Thus,
    \begin{equation}
        \pdv{\delta_{0}^{\epsilon}}{l_{03}} = -\epsilon_{1}\pdv{\beta^{\tau_{1}}_{0}}{l_{03}}-\epsilon_{2}\pdv{\beta^{\tau_{2}}_{0}}{l_{03}} = \epsilon_{0}\epsilon_{1}\epsilon_{2}\frac{S_{\Lambda}(l_{03})V_{0}}{V_{1}V_{2}},
    \end{equation}
    which is the desired identity to prove.
\end{proof}
The following proposition details how the Regge deficit angle constraint arises by reconstructing these redundant relative lengths.
\begin{proposition}[Four-point identity]
   We have, 
    \begin{equation}
        4S_{\Lambda}(l_{03})\delta_{\mathbb{R}}(D_{\Lambda}(l_{03})) = \sum_{\epsilon_{1},\epsilon_{2},\epsilon_{3} = \pm 1}\frac{S_{\Lambda}(l_{03})\delta_{2\pi}(\delta_{0}^{\epsilon})}{V_{1}V_{2}V_{3}}.\label{eq:Regge-key-identity}
    \end{equation}
    This is an equality of distributions in $l_{03}$, and is the exact two-dimensional analog of \cite[Equation (14)]{baratin_hidden_2007}, generalized to any $\Lambda$.
\end{proposition}
\begin{proof}
    At a common solution of $D_{\Lambda}(l_{03}) = 0$ and $\delta_{0}^{\epsilon} = 0$ (mod $2\pi$) we have by chain rule, Lemma \ref{lem:Gram-det-derivative}, and Lemma \ref{lem:deficit-derivative},
    \begin{equation}
        \abs{\pdv{D_{\Lambda}(l_{03})}{\delta_{0}^{\epsilon}}} = 2V_{1}V_{2}V_{3}.
    \end{equation}
    Now, because the sum over orientations $\epsilon_{1}, \epsilon_{2}, \epsilon_{3}$ counts every geometric solution twice, a standard application of the change of variables formula for delta functions gives,
    \begin{equation}
        4S_{\Lambda}(l_{03})\delta_{\mathbb{R}}(D_{\Lambda}(l_{03})) = \frac{4S_{\Lambda}(l_{03})}{2}\sum_{\epsilon_{1},\epsilon_{2},\epsilon_{3} = \pm 1}\frac{\delta_{2\pi}(\delta_{0}^{\epsilon})}{\abs{\pdv{D_{\Lambda}(l_{03})}{\delta_{0}^{\epsilon}}}} = \sum_{\epsilon_{1},\epsilon_{2},\epsilon_{3} = \pm 1}\frac{S_{\Lambda}(l_{03})\delta_{2\pi}(\delta_{0}^{\epsilon})}{V_{1}V_{2}V_{3}}
    \end{equation}
    which is exactly the claimed equality.
\end{proof}
As mentioned, by ``de-parameterizing'' the background geometry we may lift a redundant length to be dynamical, introducing a deficit angle constraint. That is, a reconstructed distance is replaced by an integration variable and a geometric compatibility constraint. In particular, for
a test function $F$,
\begin{equation}
 \sum_{\epsilon_0,\epsilon_3=\pm1}
       \frac{F(l'^{\epsilon_{0}\epsilon_{3}}_{03})}{V_0V_3}
 =4\int d l_{03}\,S_\Lambda(l_{03})
                  \delta_\R(D_\Lambda)\,F(l_{03})
 =\int d l_{03}
   \sum_{\epsilon_1,\epsilon_2,\epsilon_3=\pm1}
   \frac{S_\Lambda(l_{03})}
        {V_1V_2V_3}\,
   \delta_{2\pi}(\delta_0^\epsilon)\,F(l_{03}).
 \label{eq:missing-length-replacement}
\end{equation}
Each relative side root occurs
twice, accounting for the factor of four (see Figure \ref{fig:four-point}).
We will now use the mechanism behind this identity to reach the Regge form of \eqref{eq:relative-length-Feynman-amp} for an arbitrary triangulated disk $\Delta$, with interior vertex set $V^{0}(\Delta)\neq \emptyset$, extending Proposition \ref{prop:polygon-measure}. 

Let $\cD_\Delta$ be the domain of edge lengths for which every triangle of $\Delta$ is admissible in the sense of \S\ref{subsubsec:homogeneous-triangles}, and for every interior vertex $v$ let
\begin{equation}
    \delta_{v}^{\epsilon}(l;\Lambda):=2\pi-\sum_{\tau\ni v}\epsilon_{\tau}\beta^{\tau}_{v}(l;\Lambda)
    \label{eq:oriented-deficit}
\end{equation}
be its signed deficit angle, generalizing \eqref{eq:oriented-deficit-four}. Define
\begin{equation}
 Z_{\mathrm{LDG},\epsilon}^{\Delta,D}[\cO]
 :=
 \sum_{\{\epsilon_\tau=\pm1\}}
 \int_{\cD_\Delta}
 \left[\prod_{e\in E(\Delta)}S_\Lambda(l_e)d l_e\right]
 \left[\prod_{\tau\in T(\Delta)}\frac{1}{V_{\tau,\Lambda}}\right]
 \left[\prod_{v\in V^0(\Delta)}
            \delta_{2\pi}(\delta_v^\epsilon)\right]\cO(l,\epsilon),
 \label{eq:oriented-Regge}
\end{equation}
which is the oriented and $2\pi$-periodic version of $Z_{\mathrm{LDG}}^{\Delta, D}$ from \cite{girelli_regge_2026} (see Remark \ref{rem:int-dilaton-ori-sum}). Boundary vertices carry no deficit constraint.

Rewriting the periodic constraints with a Fourier transform, gives a local state-sum form
\begin{align}
 Z_{\mathrm{LDG},\epsilon}^{\Delta,D}[\cO]
 =\frac{1}{(2\pi)^{|V^{0}(\Delta)|}}
 \sum_{\{s_v\in\Z\}}\sum_{\{\epsilon_\tau=\pm1\}}
 \int_{\cD_\Delta}
 &\prod_{e\in E(\Delta)} S_\Lambda(l_e)d l_e\nonumber\\
 \times& \prod_{\tau\in T(\Delta)}\frac{1}{V_{\tau,\Lambda}}\,
 \mathrm{exp}\Big(i\sum_{v\in V^{0}(\Delta)}s_v\delta_v^\epsilon(l;\Lambda)\Big)
 \cO(l,\epsilon),
 \label{eq:Regge-state-sum}
\end{align}
where $s_{v}$ is playing the aforementioned role of the integer dilaton field. The sums are interpreted as Fourier representations of the distributions in \eqref{eq:oriented-Regge}, not as absolutely convergent numerical series. The integer $s_v$ is the Fourier label of the stabilizer $SO(2)$. We now arrive at our main Regge recovery theorem.

\begin{figure}[h]
\centering
\begin{tikzpicture}[scale=1.0,font=\small]
 \coordinate (p1) at (0,0); \coordinate (p2) at (2.6,-0.2); \coordinate (p3) at (3.4,1.4); \coordinate (p4) at (1.0,1.6); \coordinate (p5) at (-0.6,0.9);
 \coordinate (w) at (2.4,3.0);
 \fill[jtSage!35!white] (p1)--(p2)--(p3)--(p4)--(p5)--cycle;
 \draw[jt edge] (p1)--(p2)--(p3)--(p4)--(p5)--cycle;
 \draw[jt faint] (p1)--(p3) (p1)--(p4);
 \fill[jtSage] (p4)--(p3)--(w)--cycle;
 \draw[jt edge] (p4)--(w)--(p3);
  \node[vertex] at (p1) {};
  \node[vertex] at (p2) {};
  \node[vertex] at (p3) {};
\node[vertex] at (p5) {};  
 \node[vertex,label=below:$a$] at (p4) {};
 \node[vertex,label=below right:$b$] at (p3) {};
 \node[vertex,label=above:$w$] at (w) {};
 \node[align=center] at (1.4,-1.1) {type $(i)$: new vertex $w$,\\ new edges $(wa),(wb)$};
 \begin{scope}[xshift=7.0cm]
  \coordinate (q1) at (0,0); \coordinate (q2) at (2.6,-0.2); \coordinate (q3) at (3.4,1.4); \coordinate (qw) at (1.6,1.2); \coordinate (qa) at (0.2,2.2); \coordinate (qb) at (3.0,2.6);
  \fill[jtSage!35!white] (q1)--(q2)--(q3)--(qb)--(qw)--(qa)--cycle;
  \draw[jt edge] (q1)--(q2)--(q3)--(qb)--(qw)--(qa)--cycle;
  \draw[jt faint] (qw)--(q1) (qw)--(q2) (qw)--(q3);
  \fill[jtSage] (qa)--(qw)--(qb)--cycle;
  \draw[jt edge] (qa)--(qb);
   \node[vertex] at (q1) {};
  \node[vertex] at (q2) {};
  \node[vertex] at (q3) {};
  \node[vertex,label=left:$a$] at (qa) {};
  \node[vertex,label=right:$b$] at (qb) {};
  \node[vertex,label=below:$w$] at (qw) {};
  \node[align=center] at (1.6,-1.1) {type $(ii)$: new edge $(ab)$,\\ $w$ becomes interior};
 \end{scope}
\end{tikzpicture}
\caption{The two ways of attaching a triangle (dark green) to a triangulated disk (light green) in a shelling. The dark triangle is an ear in the first case and a closing triangle in the second.}
\label{fig:shelling}
\end{figure}
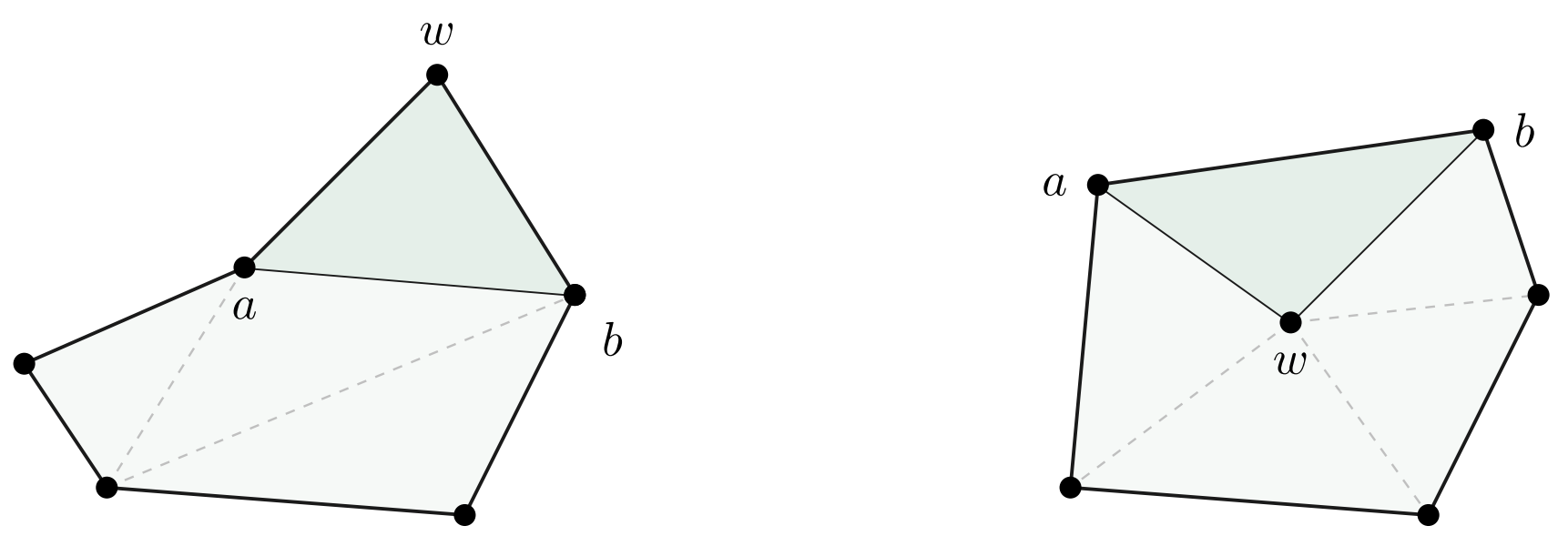

\begin{theorem}[Position measure to Regge path integral]
\label{thm:global-regge-measure}
Let $\Delta$ be any triangulation of the disk and label its $N = \abs{V(\Delta)}$ vertices by the $N$ position variables. For every diagonal-$G_{1,\kappa}$-invariant function $F$ integrable on the reduced configuration space,
\begin{equation}
 \frac{1}{\Vol(G_{1,\kappa})}\int_{X^N}\prod_{i=1}^{N}d\mu_X(x_i)\,F(\{x_i\})
 =Z_{\mathrm{LDG},\epsilon}^{\Delta,D}[F].
 \label{eq:global-position-Regge}
\end{equation}
On the right, $F$ is evaluated on the developed configuration, i.e., on the relative lengths reconstructed from $(l,\epsilon)$. Each generic labeled configuration modulo $G_{1,\kappa}$ is counted once. The assertion holds for all $\Lambda\in\R$.
\end{theorem}
\begin{proof}
We use the fact that every two-dimensional triangulated disk is shellable \cite{pachner_pl_1991, sanderson_isotopy_1957}, meaning that the triangles of $\Delta$ can be ordered as $\tau_{1},\ldots,\tau_{\abs{T(\Delta)}}$ such that every $\Delta_{n}:=\tau_{1}\cup \cdots \cup \tau_{n}\subset \Delta$ is a triangulated disk, and $\tau_{n+1}:= (wab)$ is attachable to $\Delta_{n}$ in one of two ways: ($i$) along a boundary edge $(ab)$, introducing a new vertex $w$ and the two edges $(wa), (wb)$, or ($ii$) along two consecutive boundary edges $(wa), (wb)$, introducing only the new edge $(ab)$ and turning $w$ into an interior vertex (see Figure \ref{fig:shelling}). We show by induction on $n$ that
\begin{equation}
   \frac{1}{\Vol(G_{1,\kappa})}\prod_{v\in V(\Delta_{n})}d\mu_X(x_{v})
\end{equation}
equals the measure appearing in \eqref{eq:oriented-Regge} for $\Delta_{n}$. For $n = 1$, this is simply \eqref{eq:triangle-measure}.

If the attachment is through ($i$), the new vertex $w$ contributes $d\mu_{X}(x_{w})$, which by Lemma \ref{lem:one-point-insertion} applied to the edge $(ab)$ equals,
\begin{equation}
    \sum_{\epsilon_{\tau_{n+1}} = \pm1}\frac{S_{\Lambda}(l_{wa})dl_{wa}S_{\Lambda}(l_{wb})dl_{wb}}{V_{\tau_{n+1},\Lambda}}.
\end{equation}
This is precisely the new edge factors, triangle Jacobian factor, and orientation sum with no new deficit constraint. This is the induction step of Proposition \ref{prop:polygon-measure}.

In an attachment of type ($ii$), there is no new position variable being introduced. The positions $x_{w}, x_{a}, x_{b}$ are already present and determine the value of the edge length associated to $(ab)$, together with the orientation of $\tau_{n+1} = (abw)$. We must therefore show that the new factors amount to an insertion of unity, i.e., 
\begin{equation}
 \sum_{\epsilon_{\tau_{n+1}}=\pm1}\int \frac{S_\Lambda(l_{ab})d l_{ab}}{V_{\tau_{n+1},\Lambda}}\,\delta_{2\pi}\big(\delta_w^\epsilon\big) = 1.
 \label{eq:closing-identity}
\end{equation}
Since the edge $(ab)$ is opposite to $w$ in $\tau_{n+1}$ and belongs to no other triangle containing $w$, only the angle $\beta^{\tau_{n+1}}_{w}$ in $\delta^{\epsilon}_{w}$ depends on $l_{ab}$. Differentiating the cosine law \eqref{eq:generalized-cosine-law} as in Lemma \ref{lem:one-point-insertion} gives,
\begin{equation}
    \pdv{\delta^{\epsilon}_{w}}{l_{ab}} = -\epsilon_{\tau_{n+1}}\pdv{\beta^{\tau_{n+1}}_{w}}{l_{ab}} = -\epsilon_{\tau_{n+1}}\frac{S_{\Lambda}(l_{ab})}{V_{\tau_{n+1},\Lambda}}.\label{eq:closing-jacobian}
\end{equation}
Hence, each root of $\delta^{\epsilon}_{w} = 0$ (mod $2\pi$) contributes exactly one to the left hand side of \eqref{eq:closing-identity}. For a generic configuration this fixes $\epsilon_{\tau_{n+1}}$ and $\beta^{\tau_{n+1}}_{w}$ uniquely, and the cosine law then fixes $l_{ab}$ uniquely in terms of $l_{wa}$, $l_{wb}$ and $\beta^{\tau_{n+1}}_{w}$. This proves \eqref{eq:closing-identity}.

Once $w$ is interior, no later triangle contains it, so its constraint is never modified by a subsequent attachment process and the induction process terminates with \eqref{eq:global-position-Regge}.
\end{proof}
We emphasize that the counting of variables is consistent. For any triangulated disk the Euler relation gives
\begin{equation}
 |E(\Delta)|-|V^0(\Delta)|=2N-3,
 \label{eq:stellar-count}
\end{equation}
which is indeed the number of (reduced) position variables (cf. \eqref{eq:num-redundant-lengths}). For instance, a disk obtained from one triangle by successive $1\rightarrow 3$ subdivisions has $\abs{E} = 3N-6$, $\abs{T} = 2N-5$, and $\abs{V^{0}} = N-3$.

Two important remarks are in order.

\begin{remark}[Triangulation independence]
    Since the left hand side of \eqref{eq:global-position-Regge} makes no reference to $\Delta$, the oriented Regge path integral $Z^{\Delta, D}_{\mathrm{LDG},\epsilon}$ is independent of the triangulation. That is, for any two triangulated disks $\Delta$ and $\Delta'$ on the same vertex set,
\begin{equation}
    Z^{\Delta, D}_{\mathrm{LDG},\epsilon}[F] =  Z^{\Delta', D}_{\mathrm{LDG},\epsilon}[F].\label{eq:triangulation-independence}
\end{equation}
Similarly, if $\Delta'$ is the disk obtained by subdividing one of the triangles $(123)$ of $\Delta$ by a new vertex $0$ (so that $V(\Delta') = V(\Delta)\cup \{0\}$), the left hand side acquires the factor $d\mu_{X}(x_{0})$ such that canceling the common factors one recovers an un-gauge-fixed $1\leftrightarrow 3$ identity,
\begin{equation}
 \sum_{\epsilon_{0}=\pm1}\frac{d\mu_X(x_0)}{V_0}
 =
 \sum_{\epsilon_{1},\epsilon_{2},\epsilon_{3}=\pm1}
 \frac{\prod_{i=1}^3S_\Lambda(l_{0i}) d l_{0i}}
      {V_1V_2V_3}\,
 \delta_{2\pi}(\delta_0^\epsilon).
 \label{eq:unfixed-stellar}
\end{equation}
Here $\epsilon_{0}$ is the orientation of the parent triangle $(123)$, $x_{0}$ denotes the position of the new vertex, and both sides are understood as measures on the lengths $l_{01},l_{02},l_{03}$. Note that the sum on the right cannot be restricted to the sign assignments compatible with a fixed $\epsilon_{0}$, since an assignment and its opposite may both be realized for the same $\epsilon_{0}$ (for instance $(-,-,+)$ and $(+,+,-)$ for $\epsilon_{0}=+1$, with $x_{0}$ beyond the vertex $3$ and beyond the edge $(12)$ respectively). Combined with Lemma \ref{lem:one-point-insertion}, this is nothing but \eqref{eq:missing-length-replacement}.
\end{remark}

\begin{remark}[Integer dilaton and orientation sum]\label{rem:int-dilaton-ori-sum}
    The two features of \eqref{eq:oriented-Regge} which distinguish it from the real dilaton path integral of \cite{girelli_regge_2026}, i.e., the orientation sum and the periodicity of the deficit constraint, are not additional choices, both are needed for \eqref{eq:global-position-Regge} to hold. The orientation sum is needed because the triangles of a developed configuration may be folded, and the periodicity because the developed link of an interior vertex may wind around it a number of times other than one, including zero ($w_{v}\neq1$ in \eqref{eq:angular-closure}). Dropping either restricts the integral to a set of configurations defined by $\Delta$, whereas dropping both leaves only the configurations in which every triangle of $\Delta$ is unfolded and every link winds once. The local Regge Pachner identities of \cite{girelli_regge_2026} hold on the configurations that are unfolded for both triangulations involved, such as convex quadrilaterals for the $2\leftrightarrow2$ move. Summing over orientations gives the general statement.
    
   The same two features arise when BF theory is rewritten in Regge variables (Theorems~\ref{thm:BF-recovery-from-feyn-amp} and~\ref{thm:global-regge-measure}), whereas in \cite{girelli_regge_2026} they were removed by restricting to consistently oriented triangles and to $|\delta_v|<\pi$. However, the orientation sum is readily restored there, as indicated in the proof of Lemma 4.5 of \cite{girelli_regge_2026}, the opposite orientation of each triangle supplies a separate branch of the same BF measure. By contrast, whether the periodic constraint is written with an integer dilaton or with a real dilaton and a sum over windings is a choice of representation.
\end{remark}
Combining these results with Theorem \ref{thm:BF-recovery-from-feyn-amp}, we have thus established the following theorem.

\begin{theorem}[Feynman-to-BF-to-Regge reconstruction]
\label{thm:main}
Let $\Gamma$ be a connected closed scalar graph with $N\geq3$ vertices, and let $\Delta$ be a triangulated disk with $V(\Delta) = V_{\Gamma}$ whose one-skeleton contains $\Gamma$. Then the reduced Feynman amplitude on $S^2$, $\mathbb{R}^{2}$, or $\mathbb{H}^{2}$ satisfies
\begin{equation}
 I_\Gamma^{X,\red}
 =\big\langle\cO_\Gamma^X\big\rangle_{\Delta,\BF}^{\pgf}
 =Z_{\mathrm{LDG},\epsilon}^{\Delta,D}[\cO_\Gamma^X],
 \label{eq:main}
\end{equation}
where $\cO_{\Gamma}^{X}(l) = \prod_{(ij)\in E_{\Gamma}}G_{F}(l_{ij})$. The above equalities hold for any sign of $\Lambda$, and the first equality also holds in Lorentzian signature.
\end{theorem}
\begin{proof}
The first equality is Theorem~\ref{thm:BF-recovery-from-feyn-amp}. For the second, apply Theorem~\ref{thm:global-regge-measure} to the function
\begin{equation*}
    F(\{x_{i}\}) = \prod_{(ij)\in E_{\Gamma}}G_{F}(d_{X}(x_{i},x_{j})),
\end{equation*}
which is diagonal-$G_{1,\kappa}$-invariant and, since $E_{\Gamma}\subset E(\Delta)$, depends only on the edge lengths of $\Delta$.
\end{proof}
We emphasize once more that if one restricts to configurations in $X^{N}$ in which all triangles of $\Delta$ are positively oriented and every interior vertex has winding number $w_{v} =1$ in \eqref{eq:angular-closure}, then the above considerations recover the homogeneous Regge dilaton gravity path integral $Z_{\mathrm{LDG}}^{\Delta, D}$ on the disk of \cite{girelli_regge_2026}, obtained there by integrating a real dilaton $\Phi_{v}$ at each interior vertex with the action $\sum_{v}\Phi_{v}\delta_{v}(l;\Lambda)$.

\section{Conclusion}

In this work, we have shown how Feynman diagrams for matter fields in a two-dimensional spacetime can naturally be associated with a topological theory, thereby extending the Baratin--Freidel construction from three and four dimensions to two dimensions \cite{baratin_hidden_2007,baratin_hidden_2007-1}. In particular, we have recovered both the discretized BF partition function and the two-dimensional Baratin--Freidel model from Feynman amplitudes on homogeneous backgrounds. As established in our companion paper \cite{girelli_regge_2026}, for $\Lambda \leq 0$, non-folded triangles and deficit angles vanishing exactly rather than modulo $2\pi$, these two formulations define equivalent topological invariants. A distinctive feature of the two-dimensional case is that the underlying BF theory coincides with the first order formulation of linear dilaton gravity. Our results therefore establish a direct connection between Feynman diagrams and two-dimensional linear dilaton gravity, in both the gauge theoretic and geometric formulations of the latter.

We have also emphasized why the emergence of a topological theory is natural, both from the perspective of gauging spacetime isometries and from the description of particles in terms of coadjoint orbits. These arguments are not specific to two dimensions and extend naturally to higher-dimensional homogeneous spacetimes.

An important aspect of the Baratin--Freidel models that remains to be explored in two dimensions is their relation to the $nj$-symbols associated with (quantum) group symmetries. In three dimensions the flat model is built from the $6j$-symbols of the Euclidean group \cite{baratin_hidden_2007}. By analogy, and in view of the matched pair structure of the BF gauge group used in \cite{girelli_regge_2026}, one may expect the two-dimensional model derived here to be related to the $3j$-symbols of the Euclidean group $ISO(2)$, deformed in the presence of a cosmological constant into a structure of $\kappa$-Poincar\'e type \cite{arzano_deformations_2021}. Establishing this correspondence would provide a representation-theoretic interpretation of the model and clarify the role of quantum group symmetries in its topological structure. We leave this question for future work.

\section*{Acknowledgments}\addcontentsline{toc}{section}{Acknowledgments}
FG would like to thank A. Baratin for comments and discussions. Research at the Perimeter Institute is supported in part by the Government of Canada through the Department of Innovation, Science and Economic Development and by the Province of Ontario through the Ministry of Colleges, Universities, Research Excellence and Security. We acknowledge the support of the Natural Sciences and Engineering Research Council of Canada (NSERC) through the Discovery Grant 110\_2026\_2027\_Q1\_1721 awarded to FG, and the Canadian Graduate Research Scholarship - Doctoral (CGRS D) scholarship 110\_2024\_2025\_Q2\_364 awarded to CP. We further acknowledge that many of the figures throughout this document were generated with LLM tools, and that LLMs assisted in reviewing the text for typos and overall narrative flow.

\printbibliography

@misc{bastianelli_path_2005,
  author        = {Bastianelli, Fiorenzo},
  title         = {Path Integrals in Curved Space
                   and the Worldline Formalism},
  year          = {2005},
  eprint        = {hep-th/0508205},
  archivePrefix = {arXiv},
  primaryClass  = {hep-th}
}

@article{polyakov_quantum_1981,
  author  = {Polyakov, Alexander M.},
  title   = {Quantum Geometry of Bosonic Strings},
  journal = {Physics Letters B},
  volume  = {103},
  number  = {3},
  pages   = {207--210},
  year    = {1981},
  doi     = {10.1016/0370-2693(81)90743-7}
}

@article{oeckl_generalized_2003,
  author  = {Oeckl, Robert},
  title   = {Generalized Lattice Gauge Theory, Spin Foams
             and State Sum Invariants},
  journal = {Journal of Geometry and Physics},
  volume  = {46},
  number  = {3-4},
  pages   = {308--354},
  year    = {2003},
  doi     = {10.1016/S0393-0440(02)00148-1},
  eprint  = {hep-th/0110259},
  archivePrefix = {arXiv}
}

@article{fukuma_lattice_1994,
  author  = {Fukuma, M. and Hosono, S. and Kawai, H.},
  title   = {Lattice Topological Field Theory in Two Dimensions},
  journal = {Communications in Mathematical Physics},
  volume  = {161},
  number  = {1},
  pages   = {157--175},
  year    = {1994},
  doi     = {10.1007/BF02099416},
  eprint  = {hep-th/9212154},
  archivePrefix = {arXiv}
}

@misc{ozer_holonomies_2026,
  author        = {{\"O}zer, H. T. and Filiz, Ayt{\"u}l},
  title         = {Holonomies and Boundary Symmetries in
                   Discrete {BF} Formulation of
                   Jackiw--Teitelboim Gravity},
  year          = {2026},
  eprint        = {2602.14079},
  archivePrefix = {arXiv},
  primaryClass  = {hep-th},
  doi           = {10.48550/arXiv.2602.14079}
}

@article{Birmingham:1991ty,
    author = "Birmingham, Danny and Blau, Matthias and Rakowski, Mark and Thompson, George",
    title = "{Topological field theory}",
    reportNumber = "CERN-TH-6045-91",
    doi = "10.1016/0370-1573(91)90117-5",
    journal = "Phys. Rept.",
    volume = "209",
    pages = "129--340",
    year = "1991"
}

@book{Balachandran:1983oit,
    author = "Balachandran, A. P. and Marmo, G. and Skagerstam, B. -S. and Stern, A.",
    title = "{Gauge Theories and Fibre Bundles - Applications to Particle Dynamics}",
    eprint = "1702.08910",
    archivePrefix = "arXiv",
    primaryClass = "quant-ph",
    reportNumber = "GOTEBORG-82-35",
    doi = "10.1007/3-540-12724-0_1",
    volume = "188",
    year = "1983"
}

@article{baratin_hidden_2007,
  author = {Baratin, Aristide and Freidel, Laurent},
  title = {Hidden quantum gravity in {3D} {Feynman} diagrams},
  journal = {Classical and Quantum Gravity},
  volume = {24},
  number = {8},
  pages = {1993--2026},
  year = {2007},
  month = apr,
  doi = {10.1088/0264-9381/24/8/006},
  url = {https://doi.org/10.1088/0264-9381/24/8/006},
  eprint = {gr-qc/0604016},
  archivePrefix = {arXiv},
}

@article{livine_2d_2001,
  author = {Livine, Etera R. and Perez, Alejandro and Rovelli, Carlo},
  title = {{2D} manifold-independent spinfoam theory},
  journal = {Classical and Quantum Gravity},
  volume = {20},
  number = {20},
  pages = {4425--4445},
  year = {2003},
  month = oct,
  doi = {10.1088/0264-9381/20/20/308},
  url = {https://doi.org/10.1088/0264-9381/20/20/308},
  eprint = {gr-qc/0102051},
  archivePrefix = {arXiv},
}

@article{barrett_feynman_2006,
  author = {Barrett, John W.},
  title = {{Feynman} diagrams coupled to three-dimensional quantum gravity},
  journal = {Classical and Quantum Gravity},
  volume = {23},
  number = {1},
  pages = {137--141},
  year = {2006},
  month = jan,
  doi = {10.1088/0264-9381/23/1/008},
  url = {https://doi.org/10.1088/0264-9381/23/1/008},
  eprint = {gr-qc/0502048},
  archivePrefix = {arXiv},
}

@article{baratin_hidden_2007-1,
  author = {Baratin, Aristide and Freidel, Laurent},
  title = {Hidden quantum gravity in {4D} {Feynman} diagrams: emergence of spin foams},
  journal = {Classical and Quantum Gravity},
  volume = {24},
  number = {8},
  pages = {2027--2060},
  year = {2007},
  month = apr,
  doi = {10.1088/0264-9381/24/8/007},
  url = {https://doi.org/10.1088/0264-9381/24/8/007},
  eprint = {hep-th/0611042},
  archivePrefix = {arXiv},
}

@article{oriti_spinfoam_2005,
  author = {Oriti, Daniele and Rovelli, Carlo and Speziale, Simone},
  title = {Spinfoam {2D} quantum gravity and discrete bundles},
  journal = {Classical and Quantum Gravity},
  volume = {22},
  number = {1},
  pages = {85--108},
  year = {2005},
  month = jan,
  doi = {10.1088/0264-9381/22/1/006},
  url = {https://doi.org/10.1088/0264-9381/22/1/006},
  eprint = {gr-qc/0406063},
  archivePrefix = {arXiv},
}

@article{freidel_particles_2006,
  author = {Freidel, L. and Kowalski-Glikman, J. and Starodubtsev, A.},
  title = {Particles as {Wilson} lines of the gravitational field},
  journal = {Physical Review D},
  volume = {74},
  number = {8},
  pages = {084002},
  year = {2006},
  month = oct,
  doi = {10.1103/PhysRevD.74.084002},
  url = {https://doi.org/10.1103/PhysRevD.74.084002},
  eprint = {gr-qc/0607014},
  archivePrefix = {arXiv},
}

@article{mertens_solvable_2023,
  author = {Mertens, Thomas G. and Turiaci, Gustavo J.},
  title = {Solvable models of quantum black holes: a review on {Jackiw}-{Teitelboim} gravity},
  journal = {Living Reviews in Relativity},
  volume = {26},
  number = {1},
  pages = {4},
  year = {2023},
  month = jul,
  doi = {10.1007/s41114-023-00046-1},
  url = {https://doi.org/10.1007/s41114-023-00046-1},
}

@inproceedings{goldman_locally_2010,
  author = {Goldman, William M.},
  title = {Locally homogeneous geometric manifolds},
  booktitle = {Proceedings of the International Congress of Mathematicians 2010},
  volume = {II},
  pages = {717--744},
  publisher = {Hindustan Book Agency},
  address = {New Delhi},
  year = {2011},
  month = jun,
  doi = {10.1142/9789814324359_0071},
  url = {https://doi.org/10.1142/9789814324359_0071},
  eprint = {1003.2759},
  archivePrefix = {arXiv},
  primaryClass = {math.DG},
}

@article{snobl_classification_2002,
  author = {{\v S}nobl, Libor and Hlavat{\'y}, Ladislav},
  title = {Classification of six-dimensional real {Drinfeld} doubles},
  journal = {International Journal of Modern Physics A},
  volume = {17},
  number = {28},
  pages = {4043--4067},
  year = {2002},
  month = nov,
  doi = {10.1142/S0217751X02010571},
  url = {https://doi.org/10.1142/S0217751X02010571},
  eprint = {math/0202210},
  archivePrefix = {arXiv},
}

@article{wise_symmetric_2009,
  author = {Wise, Derek K.},
  title = {Symmetric space {Cartan} connections and gravity in three and four dimensions},
  journal = {Symmetry, Integrability and Geometry: Methods and Applications},
  volume = {5},
  pages = {080},
  year = {2009},
  month = aug,
  doi = {10.3842/SIGMA.2009.080},
  url = {https://doi.org/10.3842/SIGMA.2009.080},
  eprint = {0904.1738},
  archivePrefix = {arXiv},
  primaryClass = {math.DG},
}

@article{pachner_pl_1991,
	title = {P.{L}. {Homeomorphic} {Manifolds} are {Equivalent} by {Elementary} {Shellings}},
	volume = {12},
	issn = {0195-6698},
	url = {https://www.sciencedirect.com/science/article/pii/S0195669813800807},
	doi = {10.1016/S0195-6698(13)80080-7},
	number = {2},
	journal = {European Journal of Combinatorics},
	author = {Pachner, Udo},
	month = mar,
	year = {1991},
	pages = {129--145},
}

@article{asante_quantum_2020,
    title = {Quantum geometry from higher gauge theory},
    volume = {37},
    issn = {0264-9381},
    url = {https://doi.org/10.1088/1361-6382/aba589},
    doi = {10.1088/1361-6382/aba589},
    language = {en},
    number = {20},
    journal = {Classical and Quantum Gravity},
    publisher = {IOP Publishing},
    author = {Asante, Seth K and Dittrich, Bianca and Girelli, Florian and Riello, Aldo and Tsimiklis, Panagiotis},
    month = sep,
    year = {2020},
    pages = {205001},
}

@article{ikeda_general_1993,
    author = {Ikeda, Noriaki and Izawa, Ken-Iti},
    title = {General Form of Dilaton Gravity and Nonlinear Gauge Theory},
    journal = {Progress of Theoretical Physics},
    volume = {90},
    number = {1},
    pages = {237-245},
    year = {1993},
    month = {07},
    issn = {0033-068X},
    doi = {10.1143/ptp/90.1.237},
    url = {https://doi.org/10.1143/ptp/90.1.237},
    eprint = {https://academic.oup.com/ptp/article-pdf/90/1/237/5164063/90-1-237.pdf},
}

@article{ikeda_two-dimensional_1994,
title = {Two-Dimensional Gravity and Nonlinear Gauge Theory},
journal = {Annals of Physics},
volume = {235},
number = {2},
pages = {435-464},
year = {1994},
issn = {0003-4916},
doi = {https://doi.org/10.1006/aphy.1994.1104},
url = {https://www.sciencedirect.com/science/article/pii/S0003491684711043},
author = {N. Ikeda}
}

@book{arzano_deformations_2021,
	address = {Berlin, Heidelberg},
	series = {Lecture {Notes} in {Physics}},
	title = {Deformations of {Spacetime} {Symmetries}: {Gravity}, {Group}-{Valued} {Momenta}, and {Non}-{Commutative} {Fields}},
	volume = {986},
	copyright = {https://www.springer.com/tdm},
	isbn = {978-3-662-63095-2},
	shorttitle = {Deformations of {Spacetime} {Symmetries}},
	url = {https://link.springer.com/10.1007/978-3-662-63097-6},
	doi = {10.1007/978-3-662-63097-6},
	language = {en},
	publisher = {Springer},
	author = {Arzano, Michele and Kowalski-Glikman, Jerzy},
	year = {2021},
}

@misc{girelli_regge_2026,
	title = {Regge {Dilaton} {Gravity} from {BF} {Theory}},
	url = {http://arxiv.org/abs/2610.21058},
	doi = {10.48550/arXiv.2610.21058},
	publisher = {arXiv},
	author = {Girelli, Florian and Pollack, Christopher},
	month = oct,
	year = {2026},
	note = {arXiv:2610.21058 [hep-th]},
}

@article{sanderson_isotopy_1957,
	title = {Isotopy in 3-manifolds. {I}. {Isotopic} deformations of 2-cells and 3-cells},
	volume = {8},
	issn = {0002-9939},
	url = {https://www.ams.org/journals/proc/1957-008-05/S0002-9939-1957-0090052-8/},
	doi = {10.1090/S0002-9939-1957-0090052-8},
	number = {5},
	journal = {Proceedings of the American Mathematical Society},
	author = {Sanderson, D. E.},
	year = {1957},
	pages = {912--922},
}
\end{document}